\documentclass[10pt,twoside]{article}

\usepackage[english]{babel}
\usepackage{fourier,booktabs}
\usepackage[footnotesize,bf]{caption}
\usepackage{amsfonts,amsmath,mathrsfs,tensor}

\usepackage{hyperref}

\DeclareMathAlphabet{\mathcal}{OMS}{cmsy}{m}{n}

\hypersetup{
    bookmarks=true,         % show bookmarks bar?
    unicode=false,          % non-Latin characters in Acrobat’s bookmarks
    pdftoolbar=true,        % show Acrobat’s toolbar?
    pdfmenubar=true,        % show Acrobat’s menu?
    pdffitwindow=false,     % window fit to page when opened
    pdfstartview={FitH},    % fits the width of the page to the window
    pdftitle={Interacting Scalar Fields in the Context of Effective Quantum Gravity},    % title
    pdfauthor={Artur R. Pietrykowski},     % author
    pdfsubject={Quantum field theory at frontiers},   % subject of the document
    pdfcreator={Artur R. Pietrykowski},   % creator of the document
    pdfproducer={Artur R. Pietrykowski}, % producer of the document
    pdfkeywords={quantum gravity} {triviality problem} 
                {RG group} {asymptotic safety} {Vilkovisky-DeWitt effective action}, % list of
    pdfnewwindow=true,      % links in new window
    colorlinks=true,       % false: boxed links; true: colored links
    linkcolor=blue,          % color of internal links
    citecolor=blue,        % color of links to bibliography
    filecolor=blue,      % color of file links
    urlcolor=blue           % color of external links
}

\def \a {\alpha}
\def \b {\beta}
\def \c {\chi}
\def \d {\delta}
\def \e {\epsilon}
\def \ve{\varepsilon}
\def \f {\phi}
\def \vf{\varphi}
\def \g {\gamma}
\def \et {\eta}

\def \ka {\kappa}
\def \la {\lambda}
\def \w {\omega}

\def \r {\rho}

\def \s {\sigma}

\def \th{\theta}

\def \z {\zeta}
\def \m{\mu}
\def \n{\nu}

\def \Lb{\varLambda}
\def \dl{\Delta}

\def \iggr{\mathcal{G}^{\scriptscriptstyle{-1}}}

\newcommand \gr[2]{\mathcal{G}^{{#1} , {#2}}}
\newcommand \igr[2]{\mathcal{G}^{\scriptscriptstyle{-1}}_{{#1} , {#2}}}

\def \be{\begin{equation}}
\def \ee{\end{equation}}
\def \bse{\begin{subequations}}
\def \ese{\end{subequations}}
\def \bd{\begin{displaymath}}
\def \ed{\end{displaymath}}
\def \ba{\begin{eqnarray}}
\def \ea{\end{eqnarray}}
\def \bas{\begin{eqnarray*}}
\def \eas{\end{eqnarray*}}

\def \pt{\partial}
\def \lag{\mathscr{L}}
\def \nb{\nabla}
\def \dd{\mathrm{d}}
\def \DD{\mathcal{D}}
\def \ex{\mathrm{e}}
\def \nonr{\nonumber}

\def \uf{\bar{\f}}

\def \avr#1{\langle{#1}\rangle}

\def \efa{\varGamma}

\newcommand \con[2]{\Gamma_{#2}^{#1}}

\newcommand \tr[1]{\mathrm{tr}\{{#1}\}}
\newcommand \ftr[1]{\mathrm{Tr}\left\{{#1}\right\}}

\def \mm{\int\!\dd^4\! x\ }

\def \intn#1{\int\!\dd^n\!{#1}\ {}}

\newcommand \ord[1]{\mathcal{O}({#1})}
\newcommand \cbr[1]{\left({#1}\right)}
\newcommand \sbr[1]{\left[{#1}\right]}
\newcommand \pbr[1]{\left\{{#1}\right\}}

\newcommand{\tabref}[1]{Table~\ref{tab:#1}}

\newcommand \unit[1]{\mathbb{1}_{\scriptscriptstyle({#1})}}

\numberwithin{equation}{section}

\font\lil=ptmr scaled 400

\newcommand{\preprintsize}{
      \headheight=0pt                              % header space
     \topmargin= 0cm \headsep=0cm
     \oddsidemargin= 0.3cm
      \evensidemargin= 0.3cm  % adjust left
      \textheight = 22.5truecm \textwidth=16truecm
}

\preprintsize

\title{\bf Interacting Scalar Fields in the Context of \\ Effective Quantum Gravity}

\author{Artur R. Pietrykowski\footnote{email:\href{pietrie@theor.jinr.ru}{\texttt{ pietrie@theor.jinr.ru} }  } \bigskip
\\\small{\em Bogolubov Laboratory of Theoretical Physics, JINR,}
\\\small{\em 141980 Dubna, Moscow Region, Russia}}

\date{}

\begin{document}

\maketitle

\begin{abstract}

A four dimensional scalar field theory with quartic and of higher power interactions suffers 
the triviality issue at the quantum level. This is due to 
coupling constants that, contrary to the physical expectations, seem to grow without a bound
with energy. Since this problem concerns the high energy
domain, interaction with a quantum 
gravitational field may provide natural solution to it.
In this paper we address this problem considering a scalar 
field theory with a general analytic potential having $\mathbb{Z}_2$ symmetry
and interacting with a quantum gravitational field. The dynamics of 
the latter is governed by the cosmological constant and the Einstein-Hilbert term both being
the lowest and next-to-the lowest terms of the effective theory of 
quantum gravity. Using the Vilkovisky-DeWitt method
we calculate the one loop correction to the scalar field effective action. 
We also derive the unique one loop beta functions 
for all the scalar field couplings in the MS scheme. 
We find that the leading gravitational
corrections act in the direction of asymptotic freedom. 
Moreover, assuming for both constants the Newton and the 
cosmological to have non-zero fixed point values
we find asymptotically free Halpern-Huang potentials.

\end{abstract}


\section{Introduction \label{sec1}}

The interacting scalar field theory in four spacetime dimensions is a basic constituent of perhaps the best
experimentally corroborated theory of particle physics, that is the Standard Model (SM).
In this model the Higgs particle is described by a four-component scalar field 
interacting with itself according to the quartic operator in the
interacting part of the lagrangian. It brings about a bestow of a mass upon all the 
fermions in SM as well as the part of gauge bosons obeying $SU(2)$ symmetry group
through the Higgs mechanism. A scalar field theory is also very important for cosmology where it serves to describe 
the dynamics of a very early stage of cosmological evolution, that is inflationary era. 
However, the leading order quantum corrections to the quartic coupling, that 
depend on the energy scale, revealed it to be not physically meaningful
as the ultraviolet (UV) domain is concerned. The arguments came from the one loop 
beta function to which solution is given by a relation between the momentum transfer dependent
quartic coupling $\tilde{\la}(p^2)$ and the renormalized one 
$\la_R$ at some arbitrarily chosen renormalization point.
It turns out that it increases with momentum transfer 
and at a finite value of it the effective coupling becomes infinite.  
It is usually argued that this divergence of effective coupling 
takes place at large momenta, where effective coupling is $\tilde\la(p^2)>1$ that is 
far beyond the applicability of the leading order approximation and for this 
value a sum of all orders should be taken into account.
However, investigation of $N$ component scalar field theory with $O(N)$ symmetry 
for large $N$ showed that the beta function does not depend on $N$ and 
has as the same algebraic form as in the one loop approximation \cite{Dolan:1973qd,Coleman:1974jh}, 
even though it is a function of the effective coupling rather then the renormalized one.
Since this is the non-perturbative result one concludes
that for the theory to be physically meaningful it is required
that $\la_R = 0$. This means that in the large $N$ limit a scalar field 
theory is a free field theory. Although it is not a rigorous proof
it nevertheless represents a strong premise that in general there might be no physically meaningful 
interacting scalar field theory in four spacetime dimensions. Theory with this property
is said to be trivial and nonexistence of an interacting theory is referred to 
as the triviality issue. A conjecture of triviality for scalar field theory, 
first put forward by Wilson \cite{Wilson:1971bg,Wilson:1973jj}
and examined within the functional renormalization group (RG)
\cite{Wilson:1973jj,Hasenfratz:1985dm}, has been further supported by 
large body of evidence in the Monte Carlo RG, high temperature expansion
and numerical simulations (for review see ref. \cite{Callaway:1988ya}).
An interesting discovery was made by Halpern and Huang in refs. \cite{Halpern:1994vw,Halpern:1995vf}.
They considered scalar field theory with $O(N)$ symmetry in ''Local Potential 
Approximation'' with a general potential that admits a Taylor expansion form.
In the space of all couplings,using the Wilson's RG method,
they examined a small perturbations about the free field
theory fixed point (FP) termed the Gaussian FP. 
What they found is a continuum class of nontrivial directions, 
along which the theory is asymptotically free. Potentials
along these directions are non-polynomial in fields and reveal an exponential 
growth for large scalar field values. This one loop result offers a way to evade 
the triviality issue. However, it was questioned by Morris \cite{Morris:1996nx}
as leading to wrong scaling in the large field limit as well as to singular
potentials at some value of the field in all but the Gaussian FP, even though his arguments
might implicitly assume a polynomial form of potentials \cite{Halpern:1996dh}.
Despite the doubts cast on the validity of this result, these nontrivial directions
were further investigated by many authors in various contexts 
\cite{Halpern:1997gn,Gies:2000xr,Branchina:2000jp,Huang:2011xg,Huang:2011xha}.
The triviality issue for scalar field theory and QED was recently considered
from the point of view of the Exact RG in ref. \cite{Rosten:2008ts}.
Requiring for any physical theory to have a derivative expansion
as well as demanding for it to possess a continuation 
from Euclidean to the Minkowski space it was found
that there is no physically acceptable nontrivial FPs, and the only one 
is the Gaussian FP.

The problem of triviality in an interacting scalar field theory may be rescued if
a non-Abelian gauge fields are incorporated. This phenomenon was first demonstrated in ref. \cite{Cheng:1973nv} 
in the case of Yang-Mills theory with $O(N)$ symmetry group interacting with 
scalar fields. It was found that the theory is asymptotically free in both sectors provided that $N\geq 6$.
It is therefore conceivable that if gravitational interactions
are taken into account the triviality issue may find its natural solution at the energies close to the Planck scale.
However, in the pioneering papers \cite{'tHooft:1974bx} and \cite{Goroff:1985th} it was 
revealed that the quantum field theory of gravity based on the 
Einstein-Hilbert action is non-renormalizable. Owing to Wilson's new look
at the renormalization it was realized that the renormalizable theories are
but a low energy manifestation of some underling fundamental theory that should reveal itself
in the form of a new interactions when a fundamental heavy mass threshold is approached.
This new perspective was first implemented in gravity by Donoghue \cite{Donoghue:1994dn}.
At this approach the cosmological constant and the Einstein-Hilbert term are the lowest and next-to-the lowest terms
in an energy expansion of the full theory of gravity given in the form of a covariant 
power series of interactions, each of which is formed with all possible contractions of the Riemann tensors
to a given power. The requirement of covariance makes the theory invariant with respect to
the underlying diffeomorphism symmetry.\footnote{Excellent reviews on the effective field theory may be found
in refs. \cite{Weinberg1979,Georgi:1994qn,Pich:1998xt} and 
in application to gravity in refs. \cite{Donoghue:1995cz,Burgess:2003jk} 
and the most recent \cite{Donoghue:2012zc}.} Thus
a way to study quantum effects at the low energy domain of 
quantum gravity has been opened. 
This novel point of view was utilized recently by Robinson and Wilczek \cite{Robinson:2005fj} 
to study the effects of quantum gravity on the interaction of Yang-Mills gauge fields.
It has been shown that a gravitational correction to 
running Yang-Mills coupling act in the direction of asymptotic freedom independently
of whether the symmetry group is non-Ableian or $U(1)$. 
If true, this would solve the triviality issue in the case of QED, a gauge theory with the $U(1)$ symmetry group.
The correction being quadratic in the loop momentum cut off and obtained in the momentum subtraction scheme
was questioned by many authors as gauge \cite{Pietrykowski:2006xy,Ebert:2007gf} 
and regularization dependent \cite{Toms:2007sk,Felipe:2011rs}. 
Making use of the geometric Vilkovisky-DeWitt formulation of the 
effective action and taking into account the cosmological constant
Toms \cite{Toms:2008dq,Toms:2009vd} found the gravitational correction to the Maxwell theory in the $MS$ scheme.
For positive cosmological constant the correction makes the Maxwell theory asymptotically free.
It makes the QED a nontrivial theory. A gauge independent power law gravitational correction has been found by 
Toms \cite{Toms:2010vy} through the Vilkovisky-DeWitt effective action
in conjunction with the Schwinger "proper time" method to deal with divergent loop momentum integrals. 
Although different in a form, this gravitational correction leads to the same conclusions as those drawn 
by Robinson and Wilczek \cite{Robinson:2005fj}. This result has been also derived by 
Ho et al. \cite{He:2010mt} in momentum subtraction scheme 
and corrected by Tang and Wu \cite{Tang:2011gz,Tang:2010cr} in the loop regularization scheme.
The power law correction has been criticized by may authors. Mavromatos and Ellis \cite{Ellis:2010rw} argued that
this correction is redundant and thus unphysical from the point of view of equivalence theorem. 
Anber et al. \cite{Anber:2010uj} and Anber and Donoghue \cite{Anber:2011ut}
pointed out that the power law corrections in general lead to violation of crossing symmetry and
therefore are not universal. As such, they cannot appropriately account for the quantum effects due to gravity.
The only exception from this rule is the scalar field.
Toms has recently also critically reexamined the role of the power law 
corrections \cite{Toms:2011zza} and has came to the conclusion that 
these corrections have no physical meaning. Nelsen \cite{NK2012861}
has shown in detail that the quadratic corrections depend on the gauge, even though they are 
calculated in the Vilkovisky-DeWitt formalism. The gravitational contribution 
to the running of Yang-Mills couplings has been also examined
within the asymptotic safety scenario by Daum et al. \cite{Daum:2009dn}.
Using the Euclidean and scale dependent effective action (termed the effective average action)
about the flat background metric they found a gravitational correction quadratic in IR cut off.
The correction turned out to be of the same sign as the result of Robinson and Wilczek and so
the conclusions. This result has been reexamined by Folkers et al. 
\cite{Folkerts:2011jz} where requiring for the self energy diagrams to obey a certain symmetries 
the zero result has been found. These studies imply that the status of the power law gravitational 
corrections is rather obscure from the physical point of view.
Thus the only gravitational correction contributing the running coupling involve 
the cosmological constant as found in refs. \cite{Toms:2008dq,Toms:2009vd}.

As for the scalar field a generalization of the RG methods to non-renormalizable theories
proposed by Kazakov \cite{Kazakov:1987jp} was enhanced 
and used by Barvinsky et al. \cite{Barvinsky:1993zg}, where a scalar field 
nonminimally coupled to gravity was considered. Assuming for the scalar field potential and nonminimal coupling function
to have an exponential form for large field value it was possible to solve RG equations in such a way that 
a resulting theory appeared to be asymptotically free. However, the solution 
yields unbounded from below and therefore unphysical form of the potential.
The method used in the studies was recently questioned by Steinwachs and Kamenshchik \cite{Steinwachs:2011zs}.
The effect of quantum gravity on interaction of minimally coupled scalar 
fields was also studied by Griguolo and Percacci \cite{Griguolo:1995db}
by means of the effective average action. Taking the flat background 
metric and the scalar field potential in the broken phase 
they calculated the one loop gravitational corrections to running of the 
quartic coupling and the vacuum expectation value of the scalar field.
At the high energy region the gravitational correction to the beta function
is found to be quadratic in the cut off and positive. 
This implies that the triviality problem persists. 
This result was further reexamined in ref. \cite{Percacci:2003jz} 
in the context of asymptotic safety scenario \cite{Weinberg1979} within Einstein-Hilbert truncation 
as an extension to the non-perturbative study of quantum gravity \cite{Reuter:1996cp}. 
They considered stability of the system about the
Gaussian Matter FP (GMFP) where the Newton coupling constant and the cosmological constant
both have non-zero FP values contrary to all the scalar field ones. 
Within the five coupling truncation it was found that due to the gravitational correction
the quartic operator becomes irrelevant, whereas the nonminimal operator 
$\vf^2 R$ becomes relevant. This result coincides with 
the one obtained earlier in ref. \cite{Griguolo:1995db}.
The analysis has been recently repeated and extended to arbitrary form of the potential 
including non-polynomial one as well as nonminimal coupling
function in ref. \cite{Narain:2009fy}. In the case of polynomial potentials the result
found in ref. \cite{Griguolo:1995db} was rederived. Moreover, 
investigation of stability matrix about GMFP revealed the
bidiagonal block structure of it and that each block is 
related to another by a recursion relation. Hence, 
the entire stability matrix is determined by the first diagonal
and the second diagonal block both involving solely the 
gravitational couplings. The eigenvalues were obtained only for 
the truncated potential up to mass operator with a positive real part. 
Infinite number of couplings was not considered due 
to requirements of asymptotic safety scenario, which restricts 
number of couplings at the FP. It has to be mentioned here that the results obtained by means 
of the effective average action are gauge \cite{Falkenberg:1996bq} 
and regulator \cite{Souma:2000vs} dependent. The study of influence
of quantized gravitational fields on the renormalization of a scalar field 
quartic coupling within the perturbative effective field theory
has been recently undertaken by Rodigast and Schuster \cite{Rodigast:2009zj}. From the 
Feynman diagrams they have derived the leading order of the gravitational correction to the beta function 
in the harmonic gauge that makes the scalar field theory asymptotically free. This study was extended to 
include the cosmological constant and the nonminimal coupling to gravity by MacKay and Toms \cite{Mackay:2009cf}. 
The computations have been performed within the Vilkovisky-DeWitt effective action. As a result
a gauge independent gravitational contribution to the scalar field and mass renormalizations has been found. 
The gravitational correction to the beta function for quartic coupling has not been considered there.
Finally, a very recent study of $\varphi^4$ theory in a symmetry broken phase and gravity system
within the effective approach  was undertaken by Chang at. al. \cite{Chang:2012zzo}. 
It revealed inconsistencies in a renormalization of the Higgs sector 
which is due to the gravitational corrections. This analysis, however, 
will not be addressed here.

In the present work we continue a search for the route of eschew from the problem of triviality
by encompassing quantum gravitational fluctuations. As we have seen from the 
above paragraphs this is best achieved by analysis of contributions to the RG beta functions that dictate 
the running of effective couplings. Although a form of the contributions is determined by means
of the perturbation theory it captures features that exceed the perturbative approach. 
In what follows we consider a single component scalar field theory coupled to gravity.
Since we work within effective theory we assume 
for both sectors, the scalar field and gravity, to have 
the lowest and the two derivative term.
In the case of a scalar field it corresponds to the potential and the kinetic term,
whereas in the case of the gravitational sector this corresponds
to a cosmological constant and the Einstein-Hilbert term. 
The scalar field potential is assumed to have an arbitrary, 
though analytic and $\mathbb{Z}_2$ symmetric form. Our objective is to
compute the one loop corrections to the effective action and 
derive from it the form of RG beta functions.
Since computations are performed about the flat background metric
in Euclidean space we confine the theory to be minimally 
coupled to gravity.\footnote{In a general background, however, 
terms with scalar field nonminimally coupled to gravity are required for 
reasons of renormalizability.} The flat background metric is not a solution of Einstein
equations with a cosmological constant. Therefore we perform our computations 
off the mass shell. In order to obtain gauge independent results we
employ the Vilkovisky-DeWitt geometric approach to the effective action \cite{Vilkovisky:1984st}.
Although a lack of universality of quantum corrections pointed out by Anber et al. \cite{Anber:2010uj,Anber:2011ut}
is not a concern here we use the minimal subtraction $(MS)$ scheme \cite{'tHooft:1973mm} to evade
a possible gauge dependence \cite{NK2012861}. Hence, any
quantum corrections are logarithmic in momenta. We determine the RG beta 
functions for all the non-derivative scalar field couplings along with 
the corresponding gravitational corrections. This enables
us to assess whether the gravitational corrections 
improve the high energy behaviour of the scalar field couplings.
Furthermore, owing to the Vilkovisky-DeWitt formalism 
and assuming for both gravitational couplings, the Newton and the cosmological,
to take the non-zero FP values it is possible to look for the 
asymptotically free trajectories for all the scalar field couplings.
This exploration is inspired by the Halpern-Huang discovery described in the first paragraph.
The paper is organized as follows. In section \ref{sec2:title} we introduce and motivate 
the use of the unique effective action in subsequent computations. In section \ref{sec3:title} we perform
detailed computations of the one loop correction to the Vilkovisky-DeWitt effective action. 
We compare thus obtained results with those, known in literature. 
Section \ref{sec4:title} is devoted to a study of RG equations for scalar field couplings.
The summary and conclusions are given in the final section \ref{sec5:title}.

\section{Geometric approach to the effective action \label{sec2:title}}

Standard formulation of the quantum effective action for theories with gauge symmetry
turn out to be problematic form the point of view of its applicability to the theories with
gauge symmetry. The first obstacle derives from the fact that once a gauge condition is imposed on
a variables of functional integration $\f$ to render $S_{,ij}[\f]$ invertible on the whole configuration field space
the resulting effective action, being a functional of the mean field $\uf$, 
is no longer invariant under the gauge symmetry transformation. This is
because the gauge fixing breaks also the symmetry of the mean field.
In order to keep the gauge invariance of the effective action manifest DeWitt proposed \cite{DeWitt:1967ub}
to parametrize the gauge-fixing condition for variables 
of integration $\chi^\alpha[\f]$ with some not specified external
gauge field $\varphi$ that subject background gauge transformation rules such that
the new gauge-fixing term with $\chi^\alpha[\f;\varphi]$ for quantum fields is background-gauge invariant.
However this modification worked successfully at the one loop approximation. The extension
to higher loops was proposed by 't Hooft in ref. \cite{Lopuszanski:1976vs} and further
developed by Boulware, Abbot and Hart \cite{Boulware:1980av,Abbott:1980hw,Hart:1984jy}.
The resulting effective action was gauge invariant. However, in case the equations of motions are not satisfied
it appeared to depend on the way the DeWitt's gauge fixing term is chosen.
Perhaps the easiest way to observe this dependence explicitly is to consider the one loop approximation to
the effective action. It is obtained through iterative solution of the following
equation for the background field effective action
\begin{subequations}
\begin{equation}
\label{sec2:sbfea}
\efa[\uf;\varphi] = - \log\int\DD\f\ \mathcal{M}[\f;\varphi]
\exp\pbr{- S[\f] - \tfrac{1}{2\xi}\chi^\alpha[\f;\varphi]\upsilon_{\alpha\beta}[\varphi]\chi^\beta[\f;\varphi] 
+ (\f^i-\uf^i)\frac{\d\efa[\uf;\varphi]}{\d\uf^i}}\ ,
\end{equation}
where $\xi$ is a positive real parameter. The individual quantities above are defined as follows
\begin{equation}
\label{sec2:measure}
\DD\f\ \mathcal{M}[\f;\varphi] \equiv \prod_i\dd\f^i \det\tensor{Q}{^\alpha_\beta}[\f;\varphi]\cbr{\det
\upsilon_{\alpha\beta}[\varphi]/\xi}^{\frac{1}{2}}\ ,
\end{equation}
\end{subequations}
and
\begin{equation*}
\uf^i \equiv {\avr{\f^i}}_{\! J} = \frac{\d W[J;\varphi]}{\d J_i}\ ,
\quad W[J;\varphi] = -\efa[\uf;\varphi] + J_i\uf^i\ ,
\quad \frac{\d \efa[\uf;\vf]}{\d \uf^i} = J_i\ .
\end{equation*}
The measure defined in Eq. \eqref{sec2:measure} contains the determinants of ghost operator and of
$\upsilon_{\alpha\beta}[\varphi]$ which is a non-singular matrix that 
derives from smearing with a Gaussian weight the Dirac delta
functional
inserted into the integral by the Fadeev-Popov procedure. The background field 
gauge condition has a specific form, that evades gauge fixing
of the field $\varphi$, namely
\begin{equation}
\label{sec2:defghost}
\chi^\alpha[\f;\varphi] = \chi^\alpha_{,i}[\varphi](\f^i - \varphi^i)\ ,
\quad \tensor{Q}{^\alpha_\beta}[\f;\varphi] \equiv \chi^\alpha_{,i}[\varphi]\tensor{K}{^i_\beta}[\f]\ ,
\end{equation}
where the second term defines a ghost field operator corresponding to this gauge.
The classical action $S[\f]$ is invariant under the action of the gauge group
$\mathsf{G}$ on configuration field space $\mathscr{F}$ which 
can be expressed by the infinitesimal gauge transformation, namely
\begin{equation}
\label{sec2:symmetry}
\d_\ve \f^i = \tensor{K}{^i_\alpha}[\f]\d\ve^\alpha \qquad 
\Rightarrow\qquad   S_{,i}[\f]\tensor{K}{^i_\alpha}[\f] = 0, \quad \forall \f \in
\mathscr{F}\ .
\end{equation}
In case the gauge group $\mathsf{G}$ is non-Abelian its generators $\tensor{K}{^i_\alpha}[\f]$ for
non-supersymmetric theories fulfill the following relation
\begin{equation}
\label{sec2:galgebra}
\tensor{K}{^i_{\alpha ,j}}[\f]\tensor{K}{^j_\beta}[\f] - \tensor{K}{^i_{\beta ,j}}[\f]\tensor{K}{^j_\alpha}[\f]
= \tensor{f}{^\g_{\alpha\beta}}[\f]\tensor{K}{^i_\g}[\f]\ ,
\end{equation}
where $\tensor{f}{^\g_{\alpha\beta}}[\f]$ are the structure 
functions of $\mathsf{G}$. It is assumed that the generators are
linear, i.e. $\tensor{K}{^i_{\alpha,jk}} = 0$ a 
condition that embraces the Yang-Mils as well as the gravity theory.
The structure functions in the two theories are structure constants.
The equation for the effective action in Eq. \eqref{sec2:sbfea} can be solved iteratively.
The loop expansion proceeds by changing the variable of integration $\f = \varphi+\et$ and developing
the classical action about the background field configuration $\varphi$. In the end of computations one takes the limit
$\varphi\to\uf$ the result of which is equivalent to the standard effective action but without the obstacles the original formulation
suffered. Within this limit the effective action is invariant with respect
to the transformation $\d_{\ve}\varphi = K[\varphi]\cdot\d\ve$ \cite{Boulware:1980av,Abbott:1980hw,Hart:1984jy}.
This can be proved by means of the set of identities obtained from multiple differentiation
of Eq. \eqref{sec2:symmetry} with respect to the background field
\begin{equation}
\label{sec2:wi}
S_{,i_1\dots i_n k}[\varphi]\tensor{K}{^k_\alpha}[\varphi]
+ \sum_{m = 1}^{n-1} S_{,i_1\dots i_{m-1} k\ i_{m+1}\dots i_{n-1}}[\varphi]\tensor{K}{^k_{\alpha,i_m}} = 0\ ,
\end{equation}
provided the fields $\et,\ c$ and the gauge-fixing functional 
$\c^\alpha_{,i}[\varphi]$ subject the following transformation rules
\begin{equation}
\label{sec2:gftransformrule}
\quad \d_\ve\et^i = \tensor{K}{^i_{\alpha, j}}\et^j\d\ve^\alpha ,
\quad \d_\ve\c^\alpha_{,i}[\varphi] = \cbr{\tensor{f}{^\alpha_{\beta\g}}[\varphi]\c^\g_{,i}[\varphi]
- \tensor{K}{^j_{\beta, i}}\c^\alpha_{,j}[\varphi]}\d\ve^\beta\ ,
\end{equation}
and
\begin{equation*}
\d_\ve \upsilon_{\alpha\beta}[\varphi] = \upsilon_{\alpha\beta,i}[\varphi]\tensor{K}{^i_\g}[\varphi]\d\ve^\g
=\cbr{-\tensor{f}{^\m_{\alpha\g}}[\varphi]\upsilon_{\m\beta}[\varphi]
- \tensor{f}{^\m_{\beta\g}}[\varphi]\upsilon_{\alpha\m}[\varphi]}\d\ve^\g\ .
\end{equation*}
In the case the gauge group is not compact one also has to provide some method of regularization that brings the
quantity $\tensor{f}{^\m_{\alpha\m}}[\varphi]$ to zero. Otherwise the effective
action would not be gauge invariant with respect to background
and quantum gauge transformations as well. Solving iteratively the Eq. \eqref{sec2:sbfea} up to first order
we obtain one loop effective action that takes the form
\begin{equation}
\label{sec2:1loopEA}
\efa[\varphi] = S[\varphi] + \tfrac{1}{2}\log\det\cbr{S_{,ij}[\varphi]
+ \tfrac{1}{\xi}\chi^\alpha_{,i}[\varphi] \upsilon_{\alpha\beta}[\varphi]\chi^\beta_{,j}[\varphi]} - \log\det
\tensor{Q}{^\alpha_\beta}[\varphi]\ .
\end{equation}
That this effective action depends on the gauge can be seen by considering the way
it alters if we impose the new gauge condition that differs infinitesimally from the one we had begun with.
The difference between the old and new one loop effective action amounts to
\begin{equation}
\label{sec2:gfdependence}
{\chi'}^\alpha[\f;\varphi] = \chi^\alpha[\f;\varphi] + \d\chi^\alpha[\f;\varphi]
\quad
\Rightarrow
\quad
\d_\c\efa[\varphi] =
G^{ij}[\varphi]S_{,k}[\varphi]\tensor{K}{^k_{\alpha,i}}[\varphi]
\tensor{{Q^{-1}}}{^\alpha_\beta}[\varphi]\d\chi^\beta_{,j}[\varphi]\ ,
\end{equation}
where $G$ is Green's function that is inverse to the operator defined as the argument of the first determinant
in Eq. \eqref{sec2:1loopEA}. To derive this equation we have made use of the following
identity which we will refer to as Ward identity \cite{Barvinsky:1985an}
\begin{equation}
\label{sec2:wi2}
\tfrac{1}{\xi}G^{ij}[\varphi]\chi^\alpha_{,j}[\varphi] \upsilon_{\alpha\beta}[\varphi] =
\tensor{K}{^i_\alpha}[\varphi]\tensor{{Q^{-1}}}{^\alpha_\beta}[\varphi]
+ G^{ij}S_{,k}[\varphi]\tensor{K}{^k_{\alpha,j}}[\varphi]\tensor{{Q^{-1}}}{^\alpha_\beta}[\varphi]\ .
\end{equation}
It can be obtained from the equation defining the Green's function $G$ multiplying it
by the operator $KQ^{\scriptscriptstyle -1}$, where $Q^{\scriptscriptstyle -1}$ is inverse
(Green's function) of the ghost operator \eqref{sec2:defghost} and
appropriately contracting gauge field indices.
The identity in Eq. \eqref{sec2:wi} for $n=2$ is also employed.
This result evidently shows the dependence off the mass shell on the way the gauge condition is chosen.

It was Vilkovisky who first noticed \cite{Vilkovisky:1984st} that the gauge dependence
of the effective action may be traced back to the parametrization dependence
of quantum fields. The parametrization dependence might be seen
in the term containing coupling between the difference of mean and quantum
fields and the external sources in Eq. \eqref{sec2:sbfea}. If we redefine the
variables of integration then a new variables become in general a non linear
regular local functionals $\f' = f[\f]$ of the old ones. The effective action
should be scalar w.r.t. transformations on the configuration field space which entails
$\efa[\uf] = \efa[\bar f[\f]]$ and
\begin{equation*}
\cbr{\uf^i - \f^i} \frac{\d \bar f^j[\f]}{\d \uf^i}\frac{\d \efa[\bar f]}{\d \bar f^j}
= \cbr{\bar f^i[\f] - f^i[\f]} \frac{\d \efa[\bar f]}{\d \uf^i}\ .
\end{equation*}
However, except for the specific cases, this holds for a constant matrix $\d \bar f^j[\f]\big/\d \bar \f^j$. In general
this matrix is a functional of $\f$ and this transformation rule
is valid for $\f^i$ infinitesimally close to $\bar{\f}^i$. Moreover,
in the loop expansion described above the development of the classical action about the background field
is not covariant with respect to the change of coordinates on the configuration field space $\mathscr{F}$.
Therefore the effective action is not a scalar i.e. $\efa[\uf] \neq \efa[\bar f[\f]]$.

The above arguments reveal necessity to place the formalism of the effective action
in a fully geometric setting. Therefore one regards the field configuration space $\mathscr{F}$
as a differential manifold $\mathscr{M}$ endowed with a metric
$\g$, that is $\mathscr{F} = (\mathscr{M},\g)$. Instead of using the difference of coordinates in the coupling
to the external sources which is a vector in the flat space, one uses tangent vector to
the geodesic connecting the background field with the quantum field.
This tangent vector is taken at the background field which is a point of coupling to the external sources
\begin{equation*}
\g^{ij}[\varphi]\frac{\d}{\d \varphi^j}\s[\varphi;\f]\equiv\s^i[\varphi;\f]
= -(s_2-s_1)\frac{\dd \f^i(s)}{\dd s}\Big|_{s = s_1},\quad \f^i(s_1) = \varphi^i,\ \f^i(s_2) = \f^i,
\end{equation*}
where $\s[\f,\f']$ is the half square of geodesic distance connecting the points $\f$ and $\f'$. The important
property of the quantity defined in the above equation is that 
it transforms as a vector at the background field $\varphi$
and as a scalar at the quantum field $\f$ \cite{DeWitt:1960fc}. In vicinity of the background field
the tangent vector to the geodesic has the following expansion
\begin{equation}
\label{sec2:expansion}
- \s^i[\varphi;\f] \approx \f^i - \varphi^i
+ \tfrac{1}{2}\con{i}{j k}\sbr{\g[\varphi]}(\f^j - \varphi^j)(\f^k - \varphi^k) + \dots\ ,
\end{equation}
where the symbol in front of the terms of the second order in fields denotes the Christoffel
connection built out of the metric $\g$ and its derivative to be defined below.
In flat configuration field space it vanishes so that the above quantity reduces to the difference of the coordinates
previously used to couple with the external sources.

This extension resolves the issue of a spurious quantum field coupling to the fixed external sources.
The lack of covariance that is met if one develops the classical action about the background field
in course of iterative solution for the effective action might be removed by means of the functional 
covariant derivatives replacing the usual ones. The covariant derivatives are accompanied 
with the Christoffel connection that depends on the metric $\g$ of $\mathscr{F}$.
However, the physical configuration space of the theory with 
a local gauge symmetry is a quotient space $\mathscr{F}/\mathsf{G}$.
Its elements are equivalence classes that are orbits generated by the action of the
local gauge group $\mathsf{G}$ on $\mathscr{F}$. Each member 
of the orbit of the group $\mathsf{G}$ which is a manifold itself
is enumerated by corresponding parameter $\ve^\alpha$ that 
constitutes a local coordinate on this group manifold. Thus the orbit space
$\mathscr{F}/\mathsf{G}$ along with the local gauge group $\mathsf{G}$ provide a configuration space $\mathscr{F}$
a local product structure $\mathscr{F}/\mathsf{G}\times\mathsf{G}$. From the geometric point of view this orbit
space is a submanifold endowed with an induced metric from the full configuration space metric $\g$.
Therefore the covariant derivatives on the physical configuration space should be accompanied with the Christoffel
connection evaluated on the metric of the orbit space. If we denote 
the displacement of the field coordinate in the direction of an orbit as
$\dd \f^i_{\scriptscriptstyle\parallel} = \tensor{K}{^i_\alpha}[\f]\dd\ve^\alpha$ then the one along the space of orbits can be found
from the condition $\g_{ij}[\f]\dd\f^i_{\scriptscriptstyle\perp}\dd\f^i_{\scriptscriptstyle\parallel} = 0$. Hence the metric decomposes to
\begin{equation}
\label{sec2:metrices}
\g_{ij}[\f]\dd\f^i\dd\f^j = \g^{\scriptscriptstyle\perp}_{ij}[\f]\dd\f^i_{\scriptscriptstyle\perp}\dd\f^j_{\scriptscriptstyle\perp}
+ N_{\alpha\beta}[\f]\dd\ve^\alpha\dd\ve^\beta,
\quad N_{\alpha\beta} \equiv \tensor{K}{^i_\alpha}\g_{ij}\tensor{K}{^j_\beta}\ ,
\ N_{\alpha\la}N^{\la\beta} = \d^\beta_\alpha\ ,
\end{equation}
where a tensor field $\g^{\scriptscriptstyle\perp}$ is a metric on $\mathscr{F}/\mathsf{G}$ and
$N$ is the metric on $\mathsf{G}$. The former is obtained by projection of the configuration
space metric $\g$ onto the orbit space, namely
\begin{equation}
\label{sec2:orbitspaceprojector}
\g^{\scriptscriptstyle\perp}_{ij}
\equiv P^k_i\g_{kl}P^l_j\ ,\ \g^{\scriptscriptstyle\perp}_{jk}\g_{\scriptscriptstyle\perp}^{ki} = P^i_j\ ,
\quad
P^i_j \equiv \d^i_j - \tensor{K}{^i_\alpha}N^{\alpha\beta}\tensor{K}{^k_\beta}\g_{kj}\ .
\end{equation}
Due to the terms containing $N^{\scriptscriptstyle-1}$ this metric is nonlocal.
Physical configuration space connection may be found from the condition
of compatibility of covariant derivative with the metric on $\mathscr{F}/\mathsf{G}$
that is $\nb\g^{\scriptscriptstyle\perp} = 0$ \cite{Ellicott:1990up}.
Resulting Christoffel connection constructed by means of the metric $\g^{\scriptscriptstyle\perp}$ reads
\begin{equation}
\label{sec2:defchristoffel}
\con{i}{jk}[\g^{\scriptscriptstyle\perp}] \equiv \tfrac{1}{2}
\g^{il}_{\scriptscriptstyle\perp}\cbr{\g^{\scriptscriptstyle\perp}_{jl,k}
+\g^{\scriptscriptstyle\perp}_{kl,j}-\g^{\scriptscriptstyle\perp}_{jk,l}}
= P^i_l\bar\Gamma^{l}_{jk}\ ,
\end{equation}
where the symbol on the right hand side of the above equation, that we will refer to
as the orbit space connection has the following form
\begin{equation}
\label{sec2:orbitspaceconn}
\bar\Gamma^{i}_{jk} \equiv \con{i}{jk}[\g] - 2\tensor{K}{^i_{\alpha;(j|}}N^{\alpha\beta}\tensor{K}{^l_\beta}\g_{l|k)}
+ \g_{(i| m}\tensor{K}{^m_\alpha}N^{\alpha\beta}\tensor{K}{^l_\beta}\tensor{K}{^i_{\m;l}}N^{\m\n}\tensor{K}{^n_\n}\g_{n|k)} 
+ \ldots\ ,
\quad \cbr{\tensor{K}{^i_{\alpha;j}} \equiv \nb_j[\g]\tensor{K}{^i_{\alpha}}} .
\end{equation}
The indices embraced by a curl brackets in the above formula are meant to be symmetrized.
The first term is the Christoffel connection on $\mathscr{F}$ and the second is the nonlocal
contribution that is a consequence of a projective nature of the metric on the orbit space.
As one may infer from the formula in Eq. \eqref{sec2:defchristoffel} the expression for $\bar\Gamma$
is not unique which is indicated by the ellipsis. It is 
given up to terms proportional to the generators of the gauge group.
However these terms do not contribute because any covariant derivative of the classical action
with the orbit space connection Eq. \eqref{sec2:orbitspaceconn} is orthogonal
to the symmetry directions generated by vector fields $K$ \cite{Vilkovisky:1984st}.
Moreover, due to the nonlocal part of the connection the covariant derivative
of the generator yields
\begin{equation}
\label{sec2:nablagen}
\nb_i \tensor{K}{^j_\alpha} = -\tfrac{1}{2}\tensor{K}{^j_\g}\tensor{f}{^\g_{\alpha\beta}}N^{\beta\la}\tensor{K}{^k_\la}\g_{ki} \propto
\tensor{K}{^j_\g}.
\end{equation}
The above property is crucial for the proof of gauge invariance of the effective action and of its gauge independence.
The unique or Vilkovisky-DeWitt effective action for the theories with a symmetry group
is defined as a limit in $\varphi\to\bar\f$ of the following formula
\begin{equation}
\label{sec2:VDdef}
\efa[\uf;\varphi]
\equiv -\log \int\!\!\DD\f\ \mathcal{M}[\f;\varphi]
\exp\pbr{-S[\f] - \tfrac{1}{2\xi}\chi^\alpha[\f;\varphi]\upsilon_{\alpha\beta}[\varphi]\chi^\beta[\f;\varphi]
+ \cbr{\s^i[\varphi;\bar{\f}]-\s^i[\varphi;\f]}\frac{\d\efa[\uf;\varphi]}{\d \s^i[\varphi;\bar{\f}]} } ,
\end{equation}
where
$$
\quad \mathcal{M}[\f;\varphi]\equiv \cbr{\det\g_{ij}[\f]}^{1/2}\cbr{\det{\upsilon_{\alpha\beta}[\varphi]/\xi}}^{1/2}\det
\tensor{Q}{^\alpha_\beta}[\f;\varphi]\ ,
\quad \s^i[\varphi;\bar\f] = \avr{\s^i[\varphi;\f]} .
$$
A functional fixing the gauge $\chi$ is not confined to have a specific form as in the case of
background field effective action nor must it be covariant with respect to
the background field as in Eq. \eqref{sec2:gftransformrule}. The only condition
it should satisfy is $\c[\varphi;\varphi]=0$ so that not to contribute the zeroth and first order of iterative solution to
the Eq. \eqref{sec2:VDdef}. After the limit is taken the resulting effective action has an altered form of coupling to
geodesic tangent vector field, namely
\begin{equation*}
\efa_{\!\!\!\!\mbox{\lil VD}}[\bar\f] = \lim_{\varphi\to\bar\f} \efa[\uf;\varphi]\quad
\Rightarrow
\quad \lim_{\varphi\to\bar\f}\frac{\d\efa[\uf;\varphi]}{\d \s^i[\varphi;\bar{\f}]} = 
-{C^{-1}}^j_i[\uf]{\efa_{\!\!\!\!\mbox{\lil VD}}}_{,j}[\bar\f]
\quad \cbr{C^i_j[\uf] \equiv \avr{\nb_j\s^i[\bar\f;\f]}\big|_{\f = \uf}}.
\end{equation*}
To solve the functional equation for the effective action one must first determine the form of ${C^{-1}}^i_j$,
which in turn requires the knowledge of the effective action. Thus one has to solve iteratively two coupled functional
equations. This complication is irrelevant at the one loop as ${C^{-1}}^i_j$ is a Kronecker delta and
at higher loops it may be circumvented by the method discussed by Rheban \cite{Rebhan:1987cd}.
There are two important properties that are fulfilled by the geodesic tangent vector field, namely
$$
\tensor{K}{^j_\alpha}[\varphi]\nb_j\s^i[\varphi;\f] 
= \nb_j\tensor{K}{^i_\alpha}[\varphi]\s^j[\varphi;\f]
\propto \tensor{K}{^i_\beta}[\varphi]\ ,
\quad
\frac{\d\s^i[\varphi;\f]}{\d\f^j}\tensor{K}{^j_\alpha}[\f] 
\propto \tensor{K}{^i_\beta}[\varphi].
$$
The first property follows from Eq. \eqref{sec2:nablagen}. Making 
use of these properties it may be proved that this effective
action is gauge invariant and gauge independent off the mass shell
\cite{Rebhan:1986wp,Ellicott:1990up}. Likewise the standard formulation, 
this assertion is valid provided the trace of structure constant $f^\alpha_{\ \ \ \beta \alpha}$ vanishes. 
In case of the non compact gauge groups (e.g. metric theories of gravity with the group 
of diffeomorphisms as a gauge group) this is accomplished by means of a suitable regularization. 
The most popular one is the dimensional regularization \cite{'tHooft:1973mm}. This obstacle is usually ignored 
when the ''physical'' cut-off regularization is used. 
However, it may result in the gauge parameter dependence of the final result which was recently 
exemplified by Nielsen in Einstein-Maxwell theory in ref. \cite{NK2012861}.

Iterative solution of the effective action Eq. \eqref{sec2:VDdef} proceeds in
a similar manner as in previous case. This time however the change of variables of integration is equivalent to the change
of a coordinate system in $\mathscr{F}$. Due to coupling of a tangent
geodesic vector field to the external sources in Eq. \eqref{sec2:VDdef}
the most suitable new coordinates are geodesic normal coordinates $\s^i[\varphi;\f]$.
The expansion of the classical action about the background field
is performed in an explicit covariant way, where, up to the terms needed at the one loop, it takes the form
\begin{equation*}
S[\f] = S[\varphi] - \nb_i S[\varphi]\s^{i}[\varphi;\f]
+ \tfrac{1}{2}\nb_i\nb_j S[\varphi]\s^{i}[\varphi;\f]\s^{j}[\varphi;\f] +
\ord{(\s^i)^3}.
\end{equation*}
The one loop geometric counterpart of the Eq. \eqref{sec2:1loopEA} is the Vilkovisky-DeWitt one loop effective action
\begin{equation}
\label{sec2:VD1loopea}
\efa_{\!\!\!\!\mbox{\lil VD}}[\varphi] = S[\varphi]
+ \tfrac{1}{2}\log\det\cbr{\nb_i\nb_j S[\varphi] 
+ \tfrac{1}{\xi}\c^\alpha_{,i}[\varphi]\upsilon_{\alpha\beta}[\varphi]\c^\beta_{,j}[\varphi]}
- \log\det\tensor{Q}{^\alpha_\beta}[\varphi] + \mathcal{N}[\g,\upsilon],
\end{equation}
where
$$
\mathcal{N}[\g,\upsilon,\xi] \equiv 
- \tfrac{1}{2}\log\det{\cbr{\upsilon_{\alpha\beta}[\varphi]/\xi}} 
- \tfrac{1}{2}\log\det\g_{ij}[\varphi]\ ,
$$
and the last term comes from changing variables of integration $\f\to\s$. Replacement of a functional derivative with a
covariant one in the expression in Eq. \eqref{sec2:gfdependence} and using the property in Eq. \eqref{sec2:nablagen}
shows that this effective action is independent of the gauge by virtue of Eq. \eqref{sec2:symmetry}.
The formula in Eq. \eqref{sec2:VD1loopea} involves nonlocal expressions which is due to the second 
part of orbit space connection in Eq. \eqref{sec2:orbitspaceconn}.
This non local part makes computations hardly feasible. Therefore in practice 
one chooses the orthogonal gauge \cite{Fradkin:1983nw} defined as
\begin{equation}
\label{sec2:LDgfix}
\chi^\alpha[\f;\varphi] = \upsilon^{\alpha\beta}[\varphi]
\tensor{K}{^i_\beta}[\varphi]\g_{ij}[\varphi]\s^j[\varphi;\f] = 0\ .
\end{equation}
In vicinity of $\varphi$, where according to Eq. \eqref{sec2:expansion} terms of
higher order may be neglected this gauge condition amounts
to the Landau-DeWitt gauge, provided that 
$\upsilon_{\alpha\beta}[\varphi] = c_{\alpha\beta}$ for a constant matrix $c$
and the limit $\xi \to 0$  is taken. In this gauge covariant derivative 
reduces to the local one with the Christoffel connection.
If one is able to find a new chart in which the Christoffel connection vanishes, 
then the result obtained in the unique effective action
is equivalent to that obtained in the standard background field effective action \eqref{sec2:sbfea}
\cite{Fradkin:1983nw}. In the case of gravity there are no
such coordinates, and the two results are incomparable. Within this limit the Gaussian functional with gauge
fixing term in Eq. \eqref{sec2:VDdef} shrinks to the functional Dirac delta.
The resulting effective action has as a variable of integration solely the fields $\f_{\scriptscriptstyle\perp}$ which
are nonlocal themselves. To evade this obstacle one may instead perform 
a computations with the covariant derivatives on entire $\mathscr{F}$
in the one loop effective action and in the end take the limit $\xi\to 0$. 
Thus the one loop correction to Eq. \eqref{sec2:VD1loopea}
reads
\begin{equation}
\label{sec2:VDcorr}
\efa_{\!\!\!\!\mbox{\lil VD}}^{\scriptscriptstyle (1L)}[\varphi]
= \tfrac{1}{2}\lim_{\xi\to 0}\log\det\cbr{S_{;ij}[\varphi]
+ \tfrac{1}{\xi}\g_{im}[\varphi]\tensor{K}{^m_\alpha}[\varphi]c^{\alpha\beta}\tensor{K}{^n_\beta}[\varphi]\g_{nj}[\varphi]}
- \log\det N_{\alpha\beta}[\varphi] + \ldots
\end{equation}
where we have omitted $\mathcal{N}[\g,c,\xi]$. The ghost part in this gauge amounts to the determinant 
of the metric on the group space defined in Eq. \eqref{sec2:metrices}.
In what follows we will apply the above described formalism to compute the one loop effective action
for the theory of scalar field minimally coupled to gravity.

\section{One loop effective action for gravity and scalar field system \label{sec3:title}}

Being equipped with a well established geometrical apparatus to deal
with a quantum field theories possessing a gauge symmetries we may address the question
of low energy influence of quantum gravitational degrees of freedom
carried by gravitons on scalar field defined with an arbitrary but analytic potential. Since the fundamental scale for
the theory of gravity is the Planck scale the gravitational dynamics in a low energy
limit is govern by the lowest and next to the lowest term from the infinite series of interactions
defining the effective field theory of gravity \cite{Donoghue:1994dn}. Therefore a mentioned
physical system for this energy limit is described by the following $n$-dimensional Euclidean version of the action
\begin{equation}
\label{sec3:action}
S[g,\vf] = - \frac{1}{\ka^2}\intn{x}\sqrt{g} R(g)
+ \intn{x}\sqrt{g}\cbr{\tfrac{1}{2}g^{\m\n}(\pt_\m\vf)(\pt_\n\vf)
+ U(\vf)}\ ,
\end{equation}
where $R(g)$ is the Ricci scalar and $\ka \equiv \sqrt{16\pi G}$. We assume for 
the potential of the scalar field to have the following general form
\begin{equation}
\label{sec3:origpot}
U(\vf) = \sum^\w_{n = 0} \frac{\la_{2n}}{(2n)!}\vf^{2n}\ ,
\quad \la_0 = 2\Lb/\ka^2, \ \la_2 = m^2/2,\ \la_4 = \la\ .
\end{equation}
where $\Lb$ is the cosmological constant. In what follows it will 
be convenient to redefine the scalar field in such a manner that
will enable to treat both gravitational and scalar field on equal footing. This can be attained by the following
substitution $\vf \to \vf /\ka$ which renders the scalar field dimensionless.
This redefinition produces an overall factor $1/\ka^2$ in the action i.e. $S[g,\vf]\to S[g,\vf]/\ka^2$.
Since we are interested in gravitational corrections to coupling constants
at the one-loop level we develop the action \eqref{sec3:action}, that now depends on variables
of integration $S[g^{\scriptscriptstyle q},\vf^{\scriptscriptstyle q}]$,
about the background field configuration $\vf^i = (g_{\m\n}(x),\vf(x))$
up to terms quadratic in fluctuations $\et^i = \ka(h_{\m\n}(x),\f(x))$
which is implemented by the substitution
$(g^{\scriptscriptstyle q}_{\m\n},\vf^{\scriptscriptstyle q}) = (g_{\m\n},\vf) + \ka (h_{\m\n},\f)$.
Resulting background dependent action for fluctuations  reads
\begin{equation}
\label{sec3:fluctact}
\tfrac{1}{2\ka^2} \et^i S_{,i j}[g,\vf] \et^j = \intn{x} \sqrt{g} \lag^{(2)}(x)\ ,
\quad
\lag^{(2)} = \lag^{(2)}_E + \lag^{(2)}_{\vf} + \lag^{(2)}_{\rm int}\ ,
\end{equation}
where
\begin{subequations}
\begin{eqnarray}
\label{sec3:fluctlag_gr}
\lag^{(2)}_E &=& \tfrac{1}{2} h_{\m\n}\cbr{ - \gr{\m\n}{\a\b}\Box
+ X^{\m\n ,\a\b}_{\vf} + X^{\m\n ,\a\b}_{g}}h_{\a\b} - \tfrac{1}{2}C^2_\m(h)\ ,\quad \Box \equiv g^{\m\n}\nb_\m\nb_\n\ ,
\\
\label{sec3:fluctlag_scalar}
\lag^{(2)}_{\vf} &=& \tfrac{1}{2}\f\cbr{-\Box + V''(\vf)}\f ,\quad\mbox{where}\quad V(\vf) \equiv \ka^2 U(\vf)\ ,
\\
\label{sec3:fluctlag_int}
\lag^{(2)}_{\rm int} &=& - h_{\m\n}Q^{\a|\m\n}\nb_\a \f
+ h_{\m\n} \cbr{\tfrac{1}{2}V'(\vf)g^{\m\n}}\f\ .
\end{eqnarray}
\end{subequations}
The prime in $V'(\vf)$ denotes the derivative with respect to $\vf$.
The other symbols used above are defined as follows
\begin{subequations}
\begin{eqnarray}
\label{sec3:dewitt_metric}
\gr{\m\n}{\a\b} &\equiv& \tfrac{1}{4}(g^{\m\a}g^{\n\b} + g^{\m\b}g^{\n\a} - g^{\m\n}g^{\a\b}) ,
\\
X^{\m\n ,\a\b}_{\vf} &\equiv& -\gr{\m\n}{\a\b}\sbr{\tfrac{1}{2}(\pt\vf)^2 + V(\vf)}
- \tfrac{1}{2}g^{(\m(\a}(\pt^{\b)}\vf)(\pt^{\n)}\vf) - \tfrac{1}{4}g^{\m\n}(\pt^\a\vf)(\pt^\b\vf)
- \tfrac{1}{4}g^{\a\b}(\pt^\m\vf)(\pt^\n\vf) ,
\\
\label{sec3:q}
Q^{\a|\m\n} &\equiv& g^{\a(\m}\pt^{\n)}\vf - \tfrac{1}{2} g^{\m\n}\pt^\a\vf ,
\\
\label{sec3:gauge}
C_\m(h) &\equiv& 2\gr{\a\b}{\r\s}g_{\r\m}\nb_\s h_{\a\b} = \nb^\n h_{\n\m} - \tfrac{1}{2}\pt_\m \tensor{h}{^\a_\a} \ .
\end{eqnarray}
\end{subequations}
The curl braces around indices denote the symmetrization (see Eq. \eqref{appA:symmetrization} in appendix \ref{appA}). 
The matrix $X_g$ contains a combination of Riemann tensor, Ricci tensor
and Ricci scalar, all defined on the background metric. In what follows we will, for simplicity, 
take this metric to be flat so this quantity
will vanish. The above derivation constitutes preliminary 
computations to determine a standard one loop effective action
and in consequence to find a renormalization of coupling 
constants due to interaction of scalar field with gravitons.
However the flat background metric is not a solution to the 
Einstein equations of motion derived from Eq. \eqref{sec3:action}.
From the previous section it is known that the standard 
effective action is not unique if these equations of motion are not
satisfied. Therefore in order to evade problems of gauge dependence 
related to off-shell effective action we will perform computations
by means of Vilkovisky-DeWitt geometric formalism described in previous section.

The fundamental quantity in the Vilkovisky-DeWitt formalism is a metric of
configuration field space $\mathscr{F}$ which must be a local quantity.
It is usually chosen from second order term in the expansion of a classical action about a field configuration
where it accompanies the highest-derivative d'Alembertian acting on fluctuations about this field configuration.
For the action in Eq. \eqref{sec3:action} after field redefinition, 
as described below Eq. \eqref{sec3:origpot}, the metric tensor,
as may be inferred form Eqs. \eqref{sec3:fluctact}, 
\eqref{sec3:fluctlag_gr} and \eqref{sec3:fluctlag_scalar}, takes the form
\begin{equation}
\label{sec3:configmetric}
\begin{array}{rcl}
\dd s^2 = \g_{i j}[\vf] \dd\vf^i\dd\vf^j
&=&\displaystyle\tfrac{1}{\ka^2}\intn{x}\intn{x'}\sqrt{g}\ \gr{\m\n}{\r\s}\d(x,x')\dd g_{\m\n}(x)\dd g_{\r\s}(x')
\\[6pt]
&&\displaystyle +\ \tfrac{1}{\ka^2} \intn{x} \intn{x'}\sqrt{g}\ \d(x,x')\ \dd \vf(x)\dd \vf(x')\ ,
\end{array}
\end{equation}
where $\d(x,x')$ is a density at the point $x$ and scalar at $x'$. This metric tensor may
be used to determine the orbit space connection as described in section 2.
However, in order to facilitate computations hampered by the nonlocal 
part of the orbit space connection \eqref{sec2:orbitspaceconn}
one chooses the orthogonal gauge defined in Eq. \eqref{sec2:LDgfix} in which the nonlocal part decouples.
In vicinity of the background field configuration orthogonal gauge amounts to the Landau-DeWitt gauge.
The classical action has the diffeomorphism symmetry, i.e.
\begin{equation}
\label{sec3:gauge_transforms}
\d_\ve g_{\{\m\n;x\}} = \tensor{K}{_{\{\m\n;x\} ,\{\a;y\}}}[g]\ve^{\{\a;y\}} = - 2\nb_{(\m} \ve_{\n)}(x)\ ,
\quad
\d_\ve \vf_{\{x\}} =  K_{\{x\},\{\a;y\}}[\vf]\ve^{\{\a;y\}} = - \cbr{\pt_\a\vf(x)} \ve^\a(x)\ .
\end{equation}
Thus with these generators the gauge we choose takes the form
\begin{equation}
\label{sec3:LDgauge}
\frac{\d \chi^\a}{\d \vf^i}[\vf]\et^i = c^{\a\b}\tensor{K}{^i_\b}[\vf]\g_{ij}[\vf]\et^j
= \intn{x}\!\sqrt{g}\cbr{ C^\a(h) - b (\pt^\a\vf) \f }\ .
\end{equation}
Above we introduced a parameter $b$ that in principle can assume any value.
The most popular choice is $b=0$. The Landau-DeWitt gauge requires to take $b=1$ for this parameter and we will
choose this value in the end of computations. Leaving this parameter unspecified enables us to follow the gauge dependence
of the resulting effective action. However, as it was anticipated in the end of section \ref{sec2:title}, in order to obtain the
Vilkovisky-DeWitt one loop effective action we must choose Landau-DeWitt gauge. Using the above general form of gauge the gauge-breaking
term can be reorganized to yield
\begin{equation}
\label{sec3:gfixterm}
\tfrac{1}{2\xi}\et^i \g_{ik}\tensor{K}{^k_\a}c^{\a\b}\tensor{K}{^l_\b}\g_{lj}\et^j
= \intn{x}\sqrt{g}\cbr{\tfrac{1}{2\xi}C^2_\a - \tfrac{b}{\xi}(\pt_\a h_{\m\n})Q^{\a|\m\n}\f + \tfrac{b}{2\xi}\f(\pt\vf)^2\f}\ .
\end{equation}
Due to this gauge we are left with the local part of the connection which is a Christoffel symbol
constructed by means of the metric on the full field space.
According to the definition \eqref{sec2:defchristoffel} components of the Christoffel connection for the metric \eqref{sec3:configmetric}
are
\begin{subequations}
\begin{align}
\label{sec3:congg}
\con{\{\m\n ;y\} ,\{\r\s;z\}}{\{\a\b;x\}}[\g] &= \d(x,y)\d(y,z)\sqrt{g(x)}
\cbr{-\d^{\m\n,\r\s}_{\a\b} + \tfrac{1}{4}(g^{\r\s}\d^{\m\n}_{\a\b}
+ g^{\m\n}\d^{\r\s}_{\a\b}) + \tfrac{1}{n-2}g_{\a\b}\gr{\m\n}{\r\s}}(x)\ ,
\\[5pt]
\label{sec3:conff}
\con{\{\m\n ;y\} ,\{z\}}{\{x\}}[\g] &= \tfrac{1}{4}\d(x,y)\d(y,z) \sqrt{g(x)} g^{\m\n}(x)\ ,
\\[5pt]
\label{sec3:congf}
\con{\{y\},\{z\}}{\{\a\b;x\}}[\g] &= \tfrac{1}{n-2} \d(x,y)\d(y,z)\sqrt{g(x)} g_{\a\b}(x)\  .
\end{align}
\end{subequations}
where the multi-index delta symbols are defined in Eq. \ref{appA:jedynka} of appendix. 
These connections along with the first functional derivatives of the action
that they are contracted with give the additional contribution to effective action. As it was mentioned earlier the Vilkovisky-DeWitt
formalism is defined off the mass shell. It also does not depend on the background field. Hence we may take the background metric
to be flat although it is not a solution of equation of
motion with cosmological constant. Thus in all the above formulae
we put $g_{\m\n} \to \d_{\m\n}$. Resulting first derivatives of the action with redefined fields take the form
\begin{subequations}
\begin{align}
\label{sec3:gextr}
S^{,\{\m\n;x\}} &\equiv \frac{\d S[g,\vf]}{\d g_{\m\n}(x)}\Bigg|_{g =\d}
= \d^{\m\n}\sbr{\tfrac{1}{4}(\pt\vf)^2 + \tfrac{1}{2}V(\vf)}
-\tfrac{1}{2}\pt^\m\vf\pt^\n\vf\ ,
\\
\label{sec3:fextr}
S^{,\{x\}} &\equiv \frac{\d S[g,\vf]}{\d\vf(x)} \Bigg|_{g=\d}  
= -\pt^2\vf + V'(\vf)\ , \qquad \pt^2 \equiv \d^{\m\n}\pt_\m\pt_\n\ .
\end{align}
\end{subequations}
Combining Christoffel connections from Eqs. (\ref{sec3:congg} -- \ref{sec3:congf}) 
with the above Eqs. (\ref{sec3:gextr} -- \ref{sec3:fextr})
we get the following Vilkovisky-DeWitt counterpart of the action for fluctuations in Eq. \eqref{sec3:fluctact}
supplemented with the gauge fixing term \eqref{sec3:gfixterm}, namely
\begin{equation}
\label{sec3:VDflucact}
\tfrac{1}{2}\et^i\cbr{\frac{\d^2 S}{\d\vf^i\d\vf^j}
- a \con{k}{i j} \frac{\d S}{\d \vf^k}
+ \tfrac{1}{\xi}\g_{ik}\tensor{K}{^k_\a}c^{\a\b}\tensor{K}{^l_\b}\g_{lj}}\et^j
= \intn{x} \cbr{\tilde{\lag}^{(2)}(x) + \lag^{(2)}_{\scriptscriptstyle \rm GF}(x)}\ .
\end{equation}
In the above formula we have introduced an additional parameter to be able to compare
the results between the standard one loop effective action ($a=0$)
and the Vilkovisky-DeWitt modified one ($a=1$). The quantities from
Eqs. (\ref{sec3:fluctlag_gr} -- \ref{sec3:fluctlag_int}) altered due to insertion of both connection and gauge fixing
as well as a rearrangement because of requirements of hermicity of the whole operator read
\begin{subequations}
\begin{eqnarray}
\label{sec3:lag_gr_mod}
\tilde{\lag}^{(2)} + \lag^{(2)}_{\scriptscriptstyle \rm GF}
&=& \tfrac{1}{2} h_{\m\n}\cbr{- D^{\m\n ,\a\b}(\xi,\pt)\pt^2 + \tilde{X}^{\m\n ,\a\b}}h_{\a\b}
\\
\label{sec3:lag_scalar_mod}
&& +\quad \tfrac{1}{2}\f\cbr{-\pt^2 + Y(\xi) }\f
\\
\label{sec3:lag_int_mod}
&&+\quad \tfrac{1}{2} (1-\tfrac{b}{\xi})(\pt_\a h_{\m\n})Q^{\a|\m\n}\f
-  \tfrac{1}{2} (1-\tfrac{b}{\xi})h_{\m\n}Q^{\a|\m\n}\pt_\a \f
+ \tfrac{1}{4} h_{\m\n} Z^{\m\n}(\xi) \f\ .
\end{eqnarray}
\end{subequations}
The quantities in the above operator are defined as follows
\begin{subequations}
\begin{eqnarray}
\label{sec3:operatorD}
D^{\m\n ,\a\b}(\xi,\pt) &\equiv& \gr{\m\n}{\a\b}
- \cbr{1-\tfrac{1}{\xi}}\sbr{ \d^{\m\n ,\a\b}_{\r\s}
- \tfrac{1}{2}\d^{\m\n}\d^{\a\b}_{\r\s} - \tfrac{1}{2}\d^{\a\b}\d^{\m\n}_{\r\s}
+ \tfrac{1}{4}\d^{\m\n}\d^{\a\b}\d_{\r\s} }\frac{\pt^{\r}\pt^{\s}}{\pt^2}\ ,
\\[10pt]
\label{sec3:X_mod}
\tilde{X}^{\m\n ,\a\b}&\equiv&
-\tfrac{1}{2} (1-\tfrac{a}{2}) \gr{\m\n}{\a\b}(\pt\vf)^2
+  (1 - \tfrac{a}{2}) \d^{(\m(\a}(\pt^{\b)}\vf)(\pt^{\n)}\vf)
\\\nonr
&& - \tfrac{1}{4} (1-\tfrac{a}{2})\cbr{\d^{\m\n}(\pt^\a\vf)(\pt^\b\vf)
+ \d^{\a\b}(\pt^\m\vf)(\pt^\n\vf)}
-\sbr{1-a + \tfrac{n a}{2(n-2)}}\gr{\m\n}{\a\b} V(\vf)\ ,
\\[5pt]
\label{sec3:Y_mod}
Y(\xi) &\equiv& (\tfrac{b}{\xi} - \tfrac{a}{4}) (\pt\vf)^2 + V''(\vf) - \tfrac{n a}{2(n-2)} V(\vf)\ ,
\\[5pt]
\label{sec3:z}
Z^{\m\n}(\xi) &\equiv& 2 \cbr{1 + \tfrac{b}{\xi}} \pt^\m \pt^\n \vf - \cbr{1 - a + \tfrac{b}{\xi}}\d^{\m\n}\pt^2\vf
+ ( 2 - a ) \d^{\m\n}V'(\vf)\ .
\end{eqnarray}
\end{subequations}
In order to obtain the form of the one loop correction in Eq. \eqref{sec2:VDcorr} the above formula must be completed with
the ghost Lagrangian.
In the Landau-DeWitt gauge \eqref{sec3:LDgauge} by virtue of the definition given in Eq. \eqref{sec2:metrices}
it takes the form
\begin{equation}
\label{sec3:ghost_lag}
S_{\rm ghost}[\vf,\th,\bar{\th}] = \bar\th^\a N_{\a\b}\th^\b\ ,\quad
\lag_{\rm ghost} = \bar\th^\a\cbr{-\d_{\a\b}\pt^2 - b \pt_\a\vf \pt_\b\vf}\th^\b\ .
\end{equation}
The next step that we will take in course of determining the gravitational renormalization of scalar field couplings
is the expansion of a determinant that results from a functional integration
of Eqs. (\ref{sec3:lag_gr_mod}--\ref{sec3:lag_int_mod}) and \eqref{sec3:ghost_lag} as described in previous section.

\subsection{The functional determinant and its expansion}

To find a leading quantum gravitational corrections to running of scalar coupling constants we need to compute the one loop divergences
to the kinetic term and all the vertices in the theory. Although for their derivation it is sufficient to confine oneself only to 
but a few terms that contribute to the renormalization of corresponding 
operator, we will extend computations to full scalar sector of 
the one loop effective action. This will enable us to compare the Vilkovisky-DeWitt method to 
the standard effective action results off the mass shell obtained in \cite{'tHooft:1974bx,Barvinsky:1993zg}.
Instead of using the algorithm by Barvinsky and Vilkovisky in ref. \cite{Barvinsky:1985an}
to derive the result we will use a more straightforward one that does not make use of the Ward identity given in Eq. \eqref{sec2:wi}.
It will allow us to follow the factor $1/\xi$ that should cancel in the end of computations so that the final result
would at most depend on the positive power of the gauge parameter $\xi$. This will enable us to send this parameter to zero
which is required by the Landau-DeWitt gauge. Explicit computation will allow us to verify 
the applicability of this gauge independent method to the nonrenormalizable theory first attempted in ref. \cite{Barvinsky:1985an}
in case of pure Einstein gravity and by other authors in different context in refs. \cite{Huggins:1987zw},
including recent study for the full form of the orbit space connection in 
case of the Einstein-Maxwell system undertaken in ref. \cite{NK2012861}.
The derivation of one loop effective action for nonminimal coupling of scalar field theory to gravity,
including gravitational sector, will be given elsewhere in an another context \cite{Pietrykowski}.

In the previous subsection we have determined the form of functional operator and hence, by functional integration 
over fluctuations the determinant a logarithm of which contains a 
full information about the one loop divergence structure of the scalar sector of the theory.
In order to extract this information we will expand the latter quantity 
in a series of growing number of background field dependent
vertices defined in Eqs. \eqref{sec3:q}, \eqref{sec3:X_mod}, \eqref{sec3:Y_mod} 
and \eqref{sec3:z} and keep only those terms that are divergent
in four space dimensions. The functional determinant, up to infinite constant terms reads 
\begin{eqnarray*}
\lefteqn{\tfrac{1}{2}\log\det\cbr{S_{;ij} + \tfrac{1}{\xi}\g_{ik}\tensor{K}{^k_\a} c^{\a\b}\tensor{K}{^l_\b}\g_{lj}}
- \log\det N_{\a\b}}
\\[7pt]
&=& -\log\bigg\langle \exp\pbr{ - \tfrac{1}{2} h^A \tilde{X}_{AB} h^B
-\tfrac{1}{2}\f^a Y_{ab}\f^b
+ \z h^A Q_{A a} \f^a - \z \f^a Q^{\rm T}_{a A} h^A
- \tfrac{1}{4} h^A Z_{Aa}\f^a }\bigg\rangle_{\!\!\!\scriptscriptstyle 0}
\\
&& - \log\bigg\langle\exp\pbr{- b\bar\th^\a\cbr{\pt\vf\pt\vf}_{\a\b}\th^\b}_{\a\b}\bigg\rangle_{\!\!\!\scriptscriptstyle 0} + \ldots\ ,
\qquad \cbr{i = \{A,\ a\}\quad \mbox{and}\quad A = \{\m\n;x\}\ ,\ a = \{x\}}
\end{eqnarray*}
where $\z \equiv \tfrac{1}{2}(1 - \tfrac{b}{\xi})$ and
$$
h^A Q_{A a} \f^a \equiv \intn{x} h_{\m\n}(x) Q^{\a|\m\n}(x)\pt_\a\f(x)\ ,
\quad
\f^a Q^{\rm T}_{a A} h^A \equiv \intn{x} \f(x) Q^{\a|\m\n}(x)\pt_\a h_{\m\n}(x)\ .
$$
The average is taken with the Gaussian weighting functional of the massless 
free field theory (which is indicated by subscript 0)
defined by a kinetic terms of quantum fields in Eqs. \eqref{sec3:lag_gr_mod},
\eqref{sec3:lag_scalar_mod} and \eqref{sec3:ghost_lag}. The ellipsis denote the infinite constant part.
Expanding the exponent under the functional integral, averaging
with the the Gaussian functional, making use of the Wick's theorem and finally expanding the logarithm
we arrive at the explicit form of the divergent part of $\xi$ dependent effective action.

In what follows we address the evaluation of the non-ghost as well as the ghost divergent part of the above functional determinants.
The divergent parts are extracted by means of the dimensional regularization method (DimReg), where they appear as 
a pole terms in $\e$ about the physical dimension of integrals over virtual 
particles momenta evaluated in arbitrary complex dimension $n$, i.e. for $\e = 4-n$.
The advantage of this method is that it regularizes the quadratic 
divergences to zero that would appear if momentum cut-off regularization
on virtual particles momenta was used. This solves the formal problem 
of gauge non-invariance of the functional integral measure that is met in gauge 
theories with non-compact gauge group such as the group of diffeomorphisms 
in gravity, which was mentioned in section \ref{sec2:title}. Moreover, it will allow us to extract
genuine quantum gravitational corrections that in perturbative regime contribute to 
the renormalization of the scalar field couplings as it was discussed in detail in ref. \cite{Toms:2011zza}.
Therefore in the computations we confine ourselves to the terms proportional to $1/\e$. 
The entire non ghost part of it has the following divergent contribution
\begin{subequations}
\begin{align}
\nonr
\lefteqn{\tfrac{1}{2}\log\det\cbr{S_{;ij}
+ \tfrac{1}{\xi}\g_{ik}\tensor{K}{^k_\a} c^{\a\b}\tensor{K}{^l_\b}\g_{lj}} \qquad =}
\\[10pt]
\label{sec3:firstsum}
& \tfrac{1}{2}\, \tilde{X}_{A B} G^{A B} + \tfrac{1}{2}\, {Y}_{a b} G^{a b}
- \tfrac{1}{2}\, \z^{2} {Q}_{A b} G^{A B} {Q}_{B a} G^{a b}
- \tfrac{1}{2}\, \z^{2} Q^{\rm T}_{b A} G^{A B} Q^{\rm T}_{a B} G^{a b}
\\[3pt]
\label{sec3:secondsum}
&+ \z^{2} Q^{\rm T}_{b A} G^{A B} {Q}_{B a} G^{a b} + \tfrac{1}{4}\, \z Q_{B a} G^{a b} {Z}_{b A} G^{A B}
- \tfrac{1}{4}\, \z Q^{\rm T}_{b A} G^{A B} {Z}_{B a} G^{a b}
\\[3pt]
\label{sec3:thirdsum}
& - \tfrac{1}{32}\, {Z}_{b A} G^{A B} {Z}_{B a} G^{a b}
- \tfrac{1}{4}\, \tilde{X}_{D A} G^{A B} \tilde{X}_{B C} G^{C D}
- \tfrac{1}{4}\, {Y}_{a b} G^{b c} {Y}_{c d} G^{d a}
\\[3pt]
\label{sec3:fourthsum}
& + \tfrac{1}{2}\, \z^{2} \tilde{X}_{D A} G^{A B} Q^{\rm T}_{a B} G^{a b} Q^{\rm T}_{b C} G^{C D}
- \z^{2} \tilde{X}_{D A} G^{A B} Q^{\rm T}_{a B} G^{a b} {Q}_{C b} G^{C D}
+ \tfrac{1}{2}\, \z^{2} \tilde{X}_{D A} G^{A B} {Q}_{B a} G^{a b} {Q}_{C b} G^{C D}
\\[3pt]
\label{sec3:fifthsum}
& + \tfrac{1}{2}\, \z^{2} {Y}_{d a} G^{a b} Q^{\rm T}_{b A} G^{A B} Q^{\rm T}_{c B} G^{c d}
- \z^{2} {Y}_{d a} G^{a b} Q^{\rm T}_{b A} G^{A B} {Q}_{B c} G^{c d}
+ \tfrac{1}{2}\, \z^{2} {Y}_{d a} G^{a b} {Q}_{A b} G^{A B} {Q}_{B c} G^{c d}
\\[3pt]
\label{sec3:sixthsum}
& - \tfrac{1}{4}\, \z^4 Q^{\rm T}_{a D} G^{a b} Q^{\rm T}_{b A}
G^{A B} Q^{\rm T}_{c B} G^{c d} Q^{\rm T}_{d C} G^{C D}
- \tfrac{1}{4}\, \z^{4} {Q}_{D a} G^{a b} {Q}_{A b} G^{A B}
{Q}_{B c} G^{c d} {Q}_{C d} G^{C D}
\\[3pt]
\nonr
& + \z^{4} {Q}_{D a} G^{a b} Q^{\rm T}_{b A} G^{A B} Q^{\rm T}_{c B}
G^{c d} Q^{\rm T}_{d C} G^{C D}
+ \z^{4} {Q}_{D a} G^{a b} {Q}_{A b} G^{A B} Q^{\rm T}_{c B}
G^{c d} {Q}_{C d} G^{C D}
\\[3pt] \nonr
& - \tfrac{1}{2}\, \z^{4} {Q}_{D a} G^{a b} {Q}_{A b} G^{A B}
Q^{\rm T}_{c B} G^{c d} Q^{\rm T}_{d C} G^{C D}
- \tfrac{1}{2}\, \z^{4} {Q}_{D a} G^{a b} Q^{\rm T}_{b A} G^{A B}
Q^{\rm T}_{c B} G^{c d} {Q}_{C d} G^{C D}
\\[3pt] \nonr
&- \tfrac{1}{2}\, \z^{4} {Q}_{D a} G^{a b} Q^{\rm T}_{b A} G^{A B}
{Q}_{B c} G^{c d} Q^{\rm T}_{d C} G^{C D} + \quad {\rm o.t.}
\end{align}
\end{subequations}
where ''o.t.'' indicates some other terms that do not contribute the divergent part and are omitted. 
The above symbols denote the two-point correction 
functions for graviton, scalar and ghost fields respectively defined as
$$
G^{A B} \equiv \avr{h^A h^B}_{\!\scriptscriptstyle 0}\ ,
\quad G^{a b} \equiv \avr{\f^a \f^b}_{\!\scriptscriptstyle 0}\ .
$$
Their momentum space representations take the forms respectively
\begin{equation}
\label{sec3:propagators}
G^{(h)}_{\m\n ,\r\s}(\xi,p) = \sbr{\igr{\m\n}{\r\s}
- \cbr{4(1-\xi)-4\xi\frac{M^2}{p^2}}\d^{\a\b}_{\m\n,\r\s}\frac{p_\a p_\b}{p^2} }\cbr{p^2 + M^2}^{-1}\ ,
\quad
G^{(\f)}(p) = \cbr{p^2 + M^2}^{-1}\ .
\end{equation}
$\iggr$ is the inverse of the graviton metric from 
Eq. \eqref{sec3:dewitt_metric} and $M^2$ is IR regulator.
Although there is no need for this regulator as there 
is a mass term in the theory, from the RG analysis point of view 
it is convenient to regard this mass term as a perturbation vertex. The graviton 
propagator in Eq. \eqref{sec3:propagators} owns its form to the manner 
we have introduced the IR regulator. Namely, we have modified the kinetic part 
of the operator in Eq. \eqref{sec3:lag_gr_mod} as follows 
$-h^A D_{A B}(\infty) h^B = -h_{\scriptscriptstyle\perp}^A D_{A B}(1) h_{\scriptscriptstyle\perp}^B 
\to h_{\scriptscriptstyle\perp}^A [-D_{A B}(1) + \d_{AB} M^2]h_{\scriptscriptstyle\perp}^B$,
where explicit form of $D_{AB}(\xi)$ is given in Eq. \eqref{sec3:operatorD} 
for $(A,B)=(\{x,\a\b\},\{y,\m\n\})$. $h_{\scriptscriptstyle\perp}^A\equiv P_B^A h^B$ and
$P^A_B$ is the projector on the orbit space a generic
form of which is defined in Eq. \eqref{sec2:orbitspaceprojector}.
In the end of computations we take $M\to 0$. A more difficult 
parts of algebra to be presented below were performed with the aid of
the \href{http://cadabra.phi-sci.com/}{CADABRA} software \cite{Peeters:2007wn,Peeters:CPC}.

Evaluation of the first two parts is straightforward and we find the following pole term
\begin{subequations} 
\begin{eqnarray}
\label{sec3:resultXY}
\lefteqn{\tfrac{1}{2}\, \tilde{X}_{a b} G_{\scriptscriptstyle(1)}^{a b}\big|_{\rm div} 
+ \tfrac{1}{2}\, {Y}_{a b} G_{\scriptscriptstyle(2)}^{a b}\big|_{\rm div}}
\\ \nonr
&=& -\frac{M^2}{(4\pi)^2\e}\mm\sbr{\cbr{- \tfrac{a}{4} + \tfrac{b}{\xi}}(\pt\vf)^2
+ V''(\vf) - (6 + a + 4\xi)V(\vf) }\ .
\end{eqnarray}
This term is regulator dependent and in the limit of 
vanishing $M$ there is no contribution from this part. 
The third trace from Eq. \eqref{sec3:firstsum} 
is more involved. Explicitly it takes the form
\begin{equation*}
{Q}_{A b} G^{A B} {Q}_{B a} G^{a b}
= \intn{x}\intn{x'}Q^{\la|\m\n}(x)\avr{h_{\m\n}(x)h_{\r\s}(x')} Q^{\ka|\r\s}(x')\pt_{\ka'}\pt_{\la}\avr{\f(x')\f(x)}\ .
\end{equation*}
Its evaluation can be performed in the momentum space making 
use of the formulae \eqref{sec3:propagators},
the Feynman parameters method and the averaging over 
directions. The divergences from virtual particles in the loop 
after some algebra yield the following contribution
\begin{eqnarray}
\nonr
\lefteqn{ {Q}_{A b} G^{A B} {Q}_{B a} G^{a b}\big|_{\rm div} }
\\[5pt]\nonr
&=& -\frac{2}{(4\pi)^2\e}
\Big[
\xi\z^2 M^2 \tfrac{1}{6}\ftr{Q^{\m}\unit{2}Q_{\m}} + \xi\z^2 M^2 \tfrac{1}{3}\ftr{Q_{\a}\unit{3}^{\a\b}Q_{\b}} 
- \tfrac{1}{3}\ftr{(\pt_\m Q^{\m})\iggr (\pt_\n Q^{\n})}
\\ \nonr
&& {\hskip 20pt}+ \tfrac{1}{12}\ftr{(\pt^\la Q^{\m})\iggr(\pt_\la Q_{\m})}
+ \tfrac{1}{2}(1-\xi)\ftr{(\pt_\m Q^{\m})\unit{2}(\pt_\n Q^{\n})} 
+ \tfrac{1}{6}(1-\xi)\ftr{(\pt_\a Q^{\m})\unit{3}^{\a\b}(\pt_\b Q_{\m})}
\\\nonr 
&& {\hskip 20pt}- \tfrac{2}{3}(1-\xi)\ftr{(\pt_\m Q^{\m})\unit{3}^{\a\b}(\pt_\a Q_{\b})}
- \tfrac{1}{12}(1-\xi)\ftr{(\pt_\la Q^{\m})\unit{2}(\pt^\la Q_{\m})}
- \tfrac{1}{6}(1-\xi)\ftr{(\pt_\la Q_{\a})\unit{3}^{\a\b}(\pt^\la Q_{\b})}
\Big] 
\\[10pt]
&=& \frac{M^2}{(4\pi)^2\e} \xi\z^2\mm {(\pt\vf)^2}\ ,\qquad \cbr{\ftr{AB}\equiv\intn{x}\tr{A(x)B(x)}} .
\end{eqnarray}
where $\unit{n}$ are defined in the appendix. The above shows that the
sum of traces compensate one another to yield no 
pole terms except for the regulator dependent one. 
Within the limit $M\to0$ we have no contribution 
from this part. Similar computations for the first trace 
in the Eq. \eqref{sec3:secondsum} yield the set of 
traces over discreet indices different than the above. However,
it eventually amounts to the same result. As for the first term in 
the Eq. \eqref{sec3:secondsum} its pole part reads
\begin{equation}
\label{sec3:resultQQT}
\z^{2} Q^{\rm T}_{b A} G^{A B} {Q}_{B a} G^{a b}\big|_{\rm div}
= \frac{1}{(4\pi)^2\e}\mm\Big[2 M^2 \xi\z^2 (\pt\vf)^2 + \z^2 \xi (\pt^2\vf)^2\Big]\ .
\end{equation}
Evaluation of the rest of the terms in Eqs. \eqref{sec3:secondsum} and \eqref{sec3:thirdsum}
proceeds in the same manner as sketched above. What we find for the second term of Eq. \eqref{sec3:secondsum} is
\begin{equation}
\label{sec3:resultQZ}
\begin{array}{rcl}
\tfrac{1}{4}\z {Q}_{B a} G^{a b} {Z}_{b A} G^{A B}\big|_{\rm div}
&=&\displaystyle\frac{1}{(4\pi)^2\e}\mm
\Big[-\z\cbr{\tfrac{3a}{2} + \tfrac{b}{2} + \xi (\tfrac{1}{2} - a)}(\pt^2\vf)^2
\\[5pt]
&&\displaystyle{\hskip 50pt} + \z \cbr{3 - \tfrac{3}{2}a -\xi(2-a)}(\pt\vf)^2 V''(\vf) \Big]\ ,
\end{array}
\end{equation}
and for the third term of Eq. \eqref{sec3:secondsum}
\begin{equation}
-\tfrac{1}{4}\, \z Q^{\rm T}_{b A} G^{A B} {Z}_{B a} G^{a b}\big|_{\rm div}
=\frac{1}{(4\pi)^2\e}\mm\sbr{\z\cbr{\tfrac{b}{2} + \xi(\tfrac{1}{2} 
+ \tfrac{a}{2})}(\pt^2\vf)^2 + \z \xi (-1 + \tfrac{a}{2})(\pt\vf)^2 V''(\vf) }\ .
\end{equation}
As for the traces in Eq. \eqref{sec3:thirdsum} their evaluation is straightforward and one finally finds
\begin{equation}
\label{sec3:resultZZ}
\begin{array}{rcl}
- \tfrac{1}{32} {Z}_{b A} G^{A B} {Z}_{B a} G^{a b}\big|_{\rm div}
&=& \displaystyle
\frac{1}{(4\pi)^2\e}\mm\Bigg\{ \cbr{-\tfrac{b}{2} + \tfrac{ab}{4} 
- \tfrac{b}{4\xi}(b + 3a) - \tfrac{1}{4}\xi} (\pt^2\vf)^2
\\
\ &\ &\displaystyle
+ \cbr{\tfrac{3}{2} - \tfrac{9a}{4} - \tfrac{b}{2} + \tfrac{ab}{4}
+ \tfrac{3b}{4\xi}(2-a) - \xi\cbr{ \tfrac{1}{2} - \tfrac{3a}{4}} }(\pt\vf)^2 V''(\vf)
\\
\ &\ &\displaystyle
{\hskip 90pt} + \cbr{3 - \tfrac{9}{4} a - \xi + \tfrac{3a}{4}\xi} (V'(\vf))^2
\Bigg\}\ ,
\end{array}
\end{equation}
for the first term, and the second amounts to 
\begin{equation}
\label{sec3:resultXX}
- \tfrac{1}{4} \tilde{X}_{A B} G^{B C} \tilde{X}_{C D} G^{D A} \big|_{\rm div}
= \frac{1}{(4\pi)^2\e}\mm\Bigg\{\cbr{-\tfrac{1}{2} + \tfrac{3 a}{8}}(1+\xi+\xi^2) (\pt\vf)^4
 - (3 + 2\xi^2) V^2(\vf) \Bigg\}\ ,
\end{equation}
whereas the third trace boils down to
\begin{equation}
\label{sec3:resultYY}
\begin{array}{rcl}
\displaystyle
-\tfrac{1}{4} Y_{a b} G^{b c} Y_{c d} G^{d a} \big |_{\rm div}
&=&\displaystyle 
\frac{1}{(4\pi)^2\e}\mm\Bigg [ -\tfrac{1}{2}\cbr{\tfrac{a}{4} - \tfrac{b}{\xi}}^2(\pt\vf)^4 
+ \cbr{\tfrac{a}{4} - \tfrac{b}{\xi}} (\pt\vf)^2 V''(\vf)  
\\
&&\displaystyle  
- a\cbr{\tfrac{a}{4} - \tfrac{b}{\xi}} (\pt\vf)^2 V(\vf)
- \tfrac{a}{2} V^2(\vf) - \tfrac{1}{2}  [V''(\vf)]^2
 + a V''(\vf) V(\vf)
\Bigg ]\ .
\end{array}
\end{equation}
Computation of the next few traces is slightly more complicated then those above.
Therefore we present a more detailed derivation of them.
The first trace in Eq. \eqref{sec3:fourthsum} after averaging 
over directions in momentum space and extracting of their
divergent part can be cast into the form
\begin{eqnarray*}
\nonr
\lefteqn{\tilde{X}_{A B} G^{B C} {Q}_{C a} G^{a b} {Q}_{D b} G^{D A}\big|_{\rm div} \qquad = }
\\[5pt]
&&\frac{2}{(4\pi)^2\e}
\Bigg[
\tfrac{1}{4}\delta_{\alpha \beta}\ftr{Q^\alpha
\mathcal{G}^{-1}\tilde{X} \mathcal{G}^{-1} Q^{\beta}}   
-\tfrac{1}{3}(1-\xi)\cbr{ \delta_{\alpha \beta} \delta_{\mu \nu} + 2
\delta_{\alpha \beta, \mu \nu}} \ftr{\tilde{X}
\mathcal{G}^{-1} Q^\alpha Q^\beta \unit{3}^{\mu \nu}}
\\[5pt] 
&&\tfrac{1}{12}(1-\xi)^2\cbr{ \delta_{\alpha \beta} \delta_{\mu \nu}
\delta_{\rho \sigma}
+2 \delta_{\alpha \beta} \delta_{\mu \nu,\rho \sigma}
+2 \delta_{\alpha \beta,\mu \nu} \delta_{\rho \sigma} 
+2  \delta_{\mu \nu} \delta_{\alpha \beta,\rho \sigma}
+8 \delta_{\alpha \beta, \mu \nu, \rho \sigma}}
\ftr{\tilde{X}\unit{3}^{\mu \nu} Q^\alpha Q^\beta \unit{3}^{\rho \sigma} }
\Bigg]
\\
&=& -\frac{1}{(4\pi)^2\e}\mm\sbr{\z^2\xi^2\tfrac{3}{8}(-2+a)(\pt\vf)^4
+ \z^2\xi^2 (\pt\vf)^2 V(\vf) }\ .
\end{eqnarray*}
The rest of the terms in Eq. \eqref{sec3:fourthsum} has the same form modulo sign that
comes from the different distribution of the derivatives. Accounting for the sign in front 
of the individual term the final result for the set of traces reads
\begin{equation}
\label{sec3::resultXQQ}
\begin{array}{c}
\displaystyle
\tfrac{1}{2}\, \z^{2} \tilde{X}_{A B} G^{B C} Q^{\rm T}_{a C} G^{a b} Q^{\rm T}_{b D} G^{D A}\big |_{\rm div}
- \z^{2} \tilde{X}_{A B} G^{B C} Q^{\rm T}_{a C} 
G^{a b} {Q}_{D b} G^{D A} \big |_{\rm div}
+ \tfrac{1}{2}\, \z^{2} \tilde{X}_{A B} G^{B C} {Q}_{C a} 
G^{a b} {Q}_{D b} G^{D A}\big |_{\rm div}
\\[10pt]
\displaystyle
= \qquad \frac{1}{(4\pi)^2\e}\mm\sbr{-\tfrac{3}{2}\xi^2\z^2(-2+a)(\pt\vf)^4
- 4\xi^2\z^2(\pt\vf)^2 V(\vf) }\ .
\end{array}
\end{equation}
The same remarks may be directly applied to the subsequent set of traces. Namely, 
computations of the first trace in Eq. \eqref{sec3:fifthsum} amounts to
\begin{displaymath}
Y_{a b} G^{b c} Q^T_{c A} G^{A B} Q^T_{d B} G^{d a}\big |_{\rm div} 
= \frac{2}{(4\pi)^2\e}\mm\Bigg[\cbr{b - \tfrac{a}{4}\xi}  (\pt\vf)^4 - a\xi V(\vf) (\pt\vf)^2
+ \xi V''(\vf) (\pt\vf)^2 
\Bigg]\ .
\end{displaymath}
Taking into account the different distribution of derivatives that affect the sign in front of the 
individual traces in Eq. \eqref{sec3:fifthsum} their sum yields
\begin{equation}
\label{sec3:resultYQQ}
\begin{array}{c}
\displaystyle
\tfrac{1}{2}\, \z^{2} Y_{a b} G^{b c} Q^{\rm T}_{c A} G^{A B} Q^{\rm T}_{d B} G^{d a}\big |_{\rm div}
- \z^{2} {Y}_{a b} G^{b c} Q^{\rm T}_{c A} G^{A B} {Q}_{B d} G^{d a}\big |_{\rm div}
+ \tfrac{1}{2}\, \z^{2} {Y}_{a b} G^{b c} {Q}_{A c} G^{A B} {Q}_{B d} G^{d a}\big |_{\rm div}
\\[10pt]
\displaystyle
= \qquad \frac{1}{(4\pi)^2\e}\mm
\sbr{\z^2\cbr{4 b - a\xi}  (\pt\vf)^4 - 4a\xi\z^2 V(\vf) (\pt\vf)^2
+ 4\xi\z^2 V''(\vf) (\pt\vf)^2 }\ .
\end{array}
\end{equation}
As for the last set of traces given in Eq. \eqref{sec3:sixthsum}, 
proceeding in a similar manner as in previous two sets of traces
we can confine to the first one in this equation. We find 
that the rest of them has the same abstract value, though
different sign. The first trace in Eq. \eqref{sec3:sixthsum} after some momentum space computations 
and extracting the divergent part may be cast into the following form
\begin{eqnarray*}
\nonr
\lefteqn{ {Q}_{A a} G^{A B} {Q}_{B b} G^{b c} {Q}_{C c} G^{C D} {Q}_{D d} G^{d a}\big |_{\rm div} }
\\
\nonr
&=& - \frac{2}{(4\pi)^2\e}
\Bigg[
\tfrac{1}{24}(\d_{\a\b} \d_{\g\la} + 2\d_{\a\b,\g\la})
\ftr{Q^\a Q^\b \iggr Q^\g Q^\la \iggr}
\\[5pt]
&&-(1-\xi) \tfrac{8}{4\cdot 6\cdot 8}(\d_{\a\b} \d_{\g\la} \d_{\m\n} + 2\d_{\a\b,\g\la}\d_{\m\n}
+ 2\d_{\a\b,\m\n}\d_{\g\la}+2\d_{\m\n,\g\la}\d_{\a\b} + 8\d_{\a\b,\g\la,\m\n})
\\ \nonr
&& {\hskip 100pt} \times \ftr{Q^\a Q^\b \iggr Q^\g Q^\la \unit{3}^{\m\n}}
\\[5pt]
&& + (1-\xi)^2 \tfrac{16}{4 \cdot 6 \cdot 8 \cdot 10}
(\d_{\a\b} \d_{\g\la} \d_{\m\n} \d_{\r\s} + 2 \d_{\a\b,\g\la}\d_{\m\n}\d_{\r\s}
+ 2 \d_{\a\b,\m\n}\d_{\g\la}\d_{\r\s} + 2 \d_{\a\b,\r\s}\d_{\g\la}\d_{\m\n}
\\ \nonr
&&+ 2 \d_{\g\la,\m\n}\d_{\a\b}\d_{\r\s} + 2 \d_{\g\la,\r\s}\d_{\a\b}\d_{\m\n}
+ 2 \d_{\m\n,\r\s}\d_{\a\b}\d_{\g\la} + 4 \d_{\a\b,\g\la}\d_{\m\n,\r\s} + 4 \d_{\a\b,\r\s}\d_{\g\la,\m\n}
\\ \nonr
&&
+ 4 \d_{\a\b,\m\n}\d_{\g\la,\r\s} + 8\d_{\a\b,\g\la,\m\n}\d_{\r\s} + 8\d_{\r\s,\a\b,\g\la}\d_{\m\n}
+ 8\d_{\m\n,\r\s,\a\b}\d_{\g\la} + 8\d_{\g\la,\m\n,\r\s}\d_{\a\b}
\\ \nonr
&& + 16\d_{\a\b,\g\la,\m\n,\r\s})
 \times \ftr{Q^\a Q^\b \unit{3}^{\m\n} Q^\g Q^\la \unit{3}^{\r\s} }
\Bigg]\ ,
\end{eqnarray*}
where the symbol $\d_{\a\b,\g\la,\m\n,\r\s}$ is defined in the appendix. After some algebra we obtain
\begin{equation*}
{Q}_{A a} G^{A B} {Q}_{B b} G^{b c} {Q}_{C c} G^{C D} {Q}_{D d} G^{d a}\big |_{\rm div}
= \xi^2\frac{2}{(4\pi)^2\e} \mm (\pt\vf)^4\ .
\end{equation*}
As anticipated above the rest of traces amount to the same abstract value. Taking into account the sign of each trace
contributing the sum we find the following final result for the set of traces in Eq. \eqref{sec3:sixthsum}
\begin{equation}
\label{sec3:resultQQQQ}
- \tfrac{1}{4}\, \z^4 Q^{\rm T}_{a D} G^{a b} Q^{\rm T}_{b A}
G^{A B} Q^{\rm T}_{c B} G^{c d} Q^{\rm T}_{d C} G^{C D}\big |_{\rm div} - \quad \ldots \quad 
= \quad \frac{1}{(4\pi)^2\e}\sbr{ - 8\z^4 \xi^2\mm (\pt\vf)^4}\ .
\end{equation}
\end{subequations}
The above computations are pertaining to the non ghost part. The ghost 
part of the one loop effective action may be developed as follows
\begin{equation}
\label{sec3:expghostpart}
- \log\det N_{\a\b}\big|_{\rm div}
= - b\cbr{\pt\vf\pt\vf}_{\a\b} G^{\a\b}_{\scriptscriptstyle \rm gh}
+ \tfrac{b}{2}\cbr{\pt\vf\pt\vf}_{\a\b} G^{\b\g}_{\scriptscriptstyle \rm gh}
\cbr{\pt\vf\pt\vf}_{\g\d} G^{\d\a}_{\scriptscriptstyle \rm gh}\ ,
\end{equation}
where the above symbol $G_{\scriptscriptstyle \rm gh}$ is defined along with its momentum space representation as 
\begin{equation*}
\quad G^{\a\b}_{\scriptscriptstyle \rm gh} \equiv \avr{\th^\a\bar\th^\b}_{\!\scriptscriptstyle 0}\ ,\qquad
G^{\a\b}_{\scriptscriptstyle \rm gh}(p) = \d^{\a\b}\cbr{p^2+M^2}^{-1}\ .
\end{equation*}
Evaluation of divergent part of the ghost determinant is straightforward. The first term of its 
expansion given in Eq. \eqref{sec3:expghostpart} contribute with the infrared regulator only. 
The second trace yields nonzero contribution in the $M\to0$ limit. Thus a total ghost contribution takes the form
\begin{equation}
\label{sec3:ghostresult}
-\log\det N_{\a\b}\big|_{\rm div} = \frac{1}{(4\pi)^2\e}\mm \sbr{2 b M^2 (\pt\vf)^2 + b^2 (\pt\vf)^4}\ .
\end{equation}

\subsection{The pole part of the effective action}

Assembling all the results obtained in Eqs. (\ref{sec3:resultXY}--\ref{sec3:resultQQQQ}) 
and in Eq. \eqref{sec3:ghostresult} we arrive at the final form of functional determinant. 
Retrieving the canonical dimension of the background field $\vf\to\ka\vf$ entails appropriate replacement
of the potential and its derivatives w.r.t. $\vf$, namely 
$V(\vf)\to\ka^2 U(\vf),\ V'(\vf)\to\ka U'(\vf), \ V''(\vf)\to U''(\vf)$.
Its explicit form reads
\begin{subequations}
\begin{eqnarray}
\label{sec3:det_full}
\lefteqn{\tfrac{1}{2}\log\det\cbr{S_{;ij}
+ \tfrac{1}{\xi}\g_{ik}\tensor{K}{^k_\a} c^{\a\b}\tensor{K}{^l_\b}\g_{lj}}\Big|_{\rm div} 
- \log\det N_{\a\b}\big|_{\rm div} }
\\[5pt] \nonr
&=&\frac{M^2}{(4\pi)^2\e}\mm\Bigg\{
- U''(\vf) + \sbr{\tfrac{a}{4}+ 2b + \xi - \tfrac{b}{\xi}(1-b)}\ka^2(\pt\vf)^2
+ (6 + a + 4\xi)\ka^2 U(\vf)
\Bigg\}
\\ \nonr
&& + \frac{1}{(4\pi)^2\e}\mm
\Bigg\{-\tfrac{1}{2}\cbr{U''(\vf)}^2 + A\ \ka^2(\pt^2\vf)^2 + B\ \ka^4 U^2(\vf) + C\ \ka^2\cbr{U'(\vf)}^2
+ D\ \ka^2 U''(\vf) U(\vf) 
\\ \nonr
&&\qquad+ E\ \ka^2 (\pt\vf)^2 U''(\vf) + F\ \ka^4 (\pt\vf)^2 U(\vf) + G\ \ka^4 {(\pt\vf)^4} 
\Bigg \}\ ,
\end{eqnarray}
where the coefficients in front of individual terms are defined as follows
\begin{equation}
\label{sec3:det_full_coeff}
\begin{array}{c}
\begin{array}{r@{{\ }={\ }}l@{\qquad}r @{{\ }={\ }} l}
A & - b -  \tfrac{1}{2} a b - \tfrac{3}{4} a + \tfrac{3}{4} a\xi\ , & D & a \ ,
\\[5pt]
B  & -3 - \tfrac{1}{2}a - 2\xi^2\ , & E & 3-\tfrac{11}{4}a - b - \tfrac{1}{2}ab 
+ \cbr{\tfrac{3}{2}a - 1} \xi+\tfrac{1}{\xi}b(b-1)  \ ,
\\[5pt]
C & 3 - \tfrac{9}{4} a - \xi + \tfrac{3}{4}a\xi\ , & F & - \tfrac{1}{4}a + 2 a b - b^{2} + \tfrac{1}{\xi}ab(1-b) 
+ \xi(2 b - a) - \xi^{2} \ ,
\end{array}
\\[25pt]
\begin{array}{rcl}
G &=& - \tfrac{1}{2} + \tfrac{11}{32} a  + \cbr{2- \tfrac{9}{4}b} b 
+ \tfrac{1}{2} a b(1-\tfrac{3}{4}b)
+ \cbr{-\tfrac{1}{2} + \tfrac{1}{8} a + \tfrac{1}{2}b + \tfrac{3}{4} a b} \xi
- \tfrac{1}{4} \xi^{2} 
\\[5pt]
\ & &+ \tfrac{b}{\xi}\cbr{\tfrac{1}{4}a -2b-\tfrac{1}{4}ab +2b^2}
+\tfrac{b^2}{\xi^2}\cbr{-\tfrac{1}{2}+b-\tfrac{1}{2}b^2} \ .
\end{array}
\end{array}
\end{equation}
\end{subequations}
It should be noticed that in the above result some of the coefficients of operators
depend on inverse of $\xi$ preventing us form taking the zero limit required
to obtain the Vilkovisky-DeWitt one loop correction as prescribed in Eq. \eqref{sec2:VDcorr}.
However, if we let the parameter $b$ to be such that $b^2=b$, which entails either $b=0$
or $b=1$, then all the terms with $1/\xi$ compensate one another. 
Note that this result could be obtained if there were no the configuration space connection at all
as may be checked by setting the parameter $a = 0$ in Eqs. (\ref{sec3:resultXY}--\ref{sec3:resultQQQQ}).
The presence of the gauge parameter $\xi$ and $b$ in Eq. \eqref{sec3:det_full} is a consequence of neglect of 
the nonlocal part of the orbit space connection given in Eq. \eqref{sec2:orbitspaceconn}, which
if taken into account, would also remove the terms associated with these parameters. 
In order to make up for this lack of the nonlocal part of connection we have to put the 
gauge parameter to zero as prescribed in the end of the previous section
as well as to set the parameters $a=1$ and $b=1$. This procedure 
leads to the gauge independent and gauge invariant one loop effective action. 
Thus it is another explicit example for applicability of Vilkovisky-DeWitt 
formalism to the nonrenormalizable theory, at least at the one loop level. 
Before we proceed it is interesting to compare the result in Eq. \eqref{sec3:det_full} 
for $M=0$ with those obtained by means of the standard effective action technique in various gauges. 
\begin{table}[h]
\begin{center}
\caption{Comparison of the one loop corrections to the Standard Effective Action (SEA) in various gauges parametrized 
with $\xi$ and $b$ (see Eq. \eqref{sec3:gfixterm}) with the Vilkovisky-DeWitt 
Effective Action (VDEA). $A-H$ are coefficients of the one
loop correction
given in Eq. \eqref{sec3:det_full}.}
\label{tab:table1}
\renewcommand{\arraystretch}{1.2}
\begin{tabular}{ll @{\qquad} *{7}{c}}
\toprule
method & gauge & $A$ & $B$ & $C$ & $D$ & $E$ & $F$ & $G$\\
\midrule
SEA ($a=0$)  &$\xi=0,\ b=0$ & $0$ & $-3$ & $3$ & $0$ & $3$ & $0$ & $-\tfrac{1}{2}$\\
SEA ($a=0$)  &$\xi=1,\ b=0$ & $0$ & $-5$ & $2$ & $0$ & $2$ & $-1$ & $-\tfrac{5}{4}$\\
SEA ($a=0$)  &$\xi=1,\ b=1$ & $-1$ & $-5$ & $2$ & $0$ & $1$ & $0$ & $-1$ \\
SEA ($a=0$)  &$\xi=0,\ b=1$ & $-1$ & $-3$ & $3$ & $0$ & $2$ & $-1$ & $-\tfrac{3}{4}$ \\
VDEA ($a=1$) &$\xi=0,\ b=1$ & $-\tfrac{9}{4}$ & $-\tfrac{7}{2}$ & $\tfrac{3}{4}$ & $1$ & $-\tfrac{5}{4}$ & $\tfrac{3}{4}$
& $-\tfrac{9}{32}$ \\
\bottomrule
\end{tabular}
\end{center}
\end{table}
Those results are juxtaposed in the \tabref{table1}. The first raw represents 
a set of values for the gauge parameters used in the
exact renormalization group approach to the scalar field theory non-minimally 
coupled to gravity where the beta functions for the system
has been obtained \cite{Percacci:2003jz}. A direct comparison of divergences 
is impossible though, we address the question of how to
extract the perturbative results for beta functions (see below) from the non-perturbative 
one obtained there in the last section of this paper. 
The result in the second raw of mentioned table may be confronted 
with that of refs. \cite{Gryzov:1992em}, \cite{Barvinsky:1993zg}, where
the one loop effective action for quantum gravity--nonminimally and
--minimally coupled scalar field was considered. Direct comparison 
reveals a coincidence of the abstract values of coefficients in ref.
\cite{Gryzov:1992em} and \cite{Barvinsky:1993zg} (up to a misprinted 
coefficient $B$ in the latter paper) with those displayed in the above
table. An overall sign difference comes from the different approach, 
namely Lorentzian in \cite{Barvinsky:1993zg} and Euclidean adopted in
the present paper. The gauge in the third raw of \tabref{table1} was addressed in ref. \cite{'tHooft:1974bx} (see also
\cite{Grisaru:1975ei}) where the system of scalar field minimally coupled to quantized 
gravitational field with $V(\vf)=0$ was examined. We find that the $H$ coefficient coincide with that obtained in
ref. \cite{'tHooft:1974bx}, although there is a discrepancy in the $A$ coefficient. Finally, the case in the fifth raw of
\tabref{table1} was recently considered in ref. \cite{Mackay:2009cf} for the massive scalar 
field with quartic interaction and nonminimal coupling to gravity. In order to enable this comparison 
and for the sake of the further discussion we adopt the potential in the form given in Eq. \eqref{sec3:origpot}
and confine our considerations up to $\vf^4$ and $(\pt\vf)^2$ terms. 
Reinstating the original definition of the scalar field which is
implemented by replacing $\vf\to\ka\vf$ the resulting Vilkovisky-DeWitt one loop effective action reads
\begin{subequations}
\begin{equation}
\label{sec3:VD1loopEA_result}
\efa_{\!\!\!\!\mbox{\lil VD}}^{\scriptscriptstyle (1L)}[\vf] 
= - \tfrac{1}{\e} \mm\Big[ A(\pt^2\vf)^2 
+ z^{\scriptscriptstyle(1|1L)}_\vf \tfrac{1}{2}(\pt\vf)^2 
+z^{\scriptscriptstyle(1|1L)}_{\vf^2} \tfrac{1}{2} m^2 \vf^2 
+z^{\scriptscriptstyle(1|1L)}_{\vf^4}\tfrac{\la}{4!} \vf^4 \Big] ,
\end{equation}
where
\begin{equation}
\label{sec3:VD1loopEA_result_coeff}
\begin{array}{r@{{\ }={\ }}l@{\qquad}r @{{\ }={\ }} l}
(4\pi)^2 A  & \tfrac{9}{4}\ka^2\ , & (4\pi)^2 z^{\scriptscriptstyle(1|1L)}_{\vf^2} & 14 \ka^2\Lb + \cbr{1  - 2\Lb m^{-2}}\la -
\tfrac{5}{2}\ka^2 m^2\ , 
\\[5pt]
(4\pi)^2 z^{\scriptscriptstyle(1|1L)}_\vf & -3\ka^2\Lb + \tfrac{5}{2}\ka^2 m^2\ ,
& (4\pi)^2 z^{\scriptscriptstyle(1|1L)}_{\vf^4} & 3\la + (14\Lb - 13 m^2)\ka^2 + 21 m^4\ka^4\la^{-1}\ .
\end{array}
\end{equation}
\end{subequations}
From the comparison of the above coefficients with ref. \cite{Mackay:2009cf} in the case of vanishing nonminimal coupling 
and taking into account the different definition of gravitational coupling (the relation is $\ka^2 = \tilde \ka^2/2$, 
where LHS denotes the definition given below Eq. \eqref{sec3:action}), aside of a misprint in 
$\Lb$ accompanied factor in Eq. ($37$) of this paper\footnote{This $\Lb$ accompanied factor in ref. \cite{Mackay:2009cf}, 
according to the definition of $B$ in Eq. ($35$), in the limit $\w ,\n \to 1$ 
and $\a,\xi_{\rm nmc} \to 0$ is equal to $-3/8$ instead of $-1/2$ given there in Eq. (37).}, 
we find a full agreement up to the term $\vf^2$. The coefficient of the quartic coupling is missing there.

The one loop correction to the effective action given in Eq. \eqref{sec3:det_full_coeff} 
is related to the one loop counter term by the equation 
$\dl S|_{\scriptscriptstyle 1L} = - \efa_{\!\!\!\!\mbox{\lil VD}}^{\scriptscriptstyle (1L)}$. 
If we take the limit $M\to0$ and adopt the potential to have the form 
given in Eq. \eqref{sec3:origpot} then the bare action reads 
$$
S_B[\vf] = S[\vf] + \dl S[\vf]\ ,
$$
where the counter term takes the form
\begin{eqnarray}
\label{sec7:counterterm}
\dl S[\vf] &=& \mm\Big[Z^{\scriptscriptstyle(1)}_\vf\tfrac{1}{2}(\pt\vf)^2 
+ \sum^\w_{n=0}Z^{\scriptscriptstyle(1)}_{\vf^{2n}}\tfrac{1}{(2n)!}\la_{2n}\vf^{2n} 
\\
\nonr
&& {\hskip 60pt}+\sum^\w_{n = 1}Z^{\scriptscriptstyle(1)}_{(\pt\vf)^2\vf^{2n}}(\pt\vf)^2\vf^{2n} 
+ Z^{\scriptscriptstyle(1)}_{(\pt\vf)^4}(\pt\vf)^4\Big]\ .
\end{eqnarray}
The coefficients in front of operators are related to the corresponding renormalization constants
by the equation
\begin{equation}
\label{sec3:renorm_consts}
Z_{\mathcal{O}}(g,\e) = 1 + \sum_{\n\geq 1}Z^{\scriptscriptstyle(\n)}_{\mathcal{O}}(g)\e^{-\n} \ ,
\quad Z^{\scriptscriptstyle(\n)}_{\mathcal{O}}(g) = \sum_{r\geq 1}z^{\scriptscriptstyle(\n|rL)}_{\mathcal{O}}(g)\ ,
\quad \mathcal{O} = \{\vf, \vf^{2n}, (\pt\vf)^2\vf^{2n}, (\pt\vf)^4\}\ .
\end{equation}
The form of the first two one loop renormalization constants 
may be inferred from the Eq. \eqref{sec3:det_full_coeff} and read 
\begin{subequations}
\begin{equation}
\label{sec3:renorm_const_field}
(4\pi)^2 z^{\scriptscriptstyle(1|1L)}_\vf = - 2 E \ka^2 \la_2 - 2 F \ka^4\la_0 \ ,
\end{equation}
and
\begin{equation}
\label{sec3:renorm_const_vertices}
(4\pi)^2 z^{\scriptscriptstyle(1|1L)}_{\vf^{2n}}
= \frac{1}{\la_{2n}}\sum_{k=0}^n \binom{2n}{2k} 
\Bigg\{
\tfrac{1}{2} \la_{2(k+1)}\la_{2(n-k+1)}
-\cbr{C \frac{2(n-k)}{2k+1} + D}\ka^2\la_{2(k+1)}\la_{2(n-k)} 
- B \ka^4 \la_{2k}\la_{2(n-k)}
\Bigg\}\ ,
\end{equation}
\end{subequations}
respectively. The rest of the one loop renormalization constants can be readily inferred from 
the mentioned formula. However, as they are not to be further utilized we will keep them implicit. 
Having evaluated the form of the counter term and one loop renormalization constants we can derive 
out of it equations for running couplings in the theory under considerations.

\section{Running scalar field couplings in the MS scheme \label{sec4:title}}

Let us address the question of how couplings in the action \eqref{sec3:action}
with the general form of the potential given in
Eq. \eqref{sec3:origpot} change with respect to the energy scale.
In a full effective theory the set of couplings consists of derivative and non-derivative ones.
Since we have restricted the effective action to the lowest energy 
terms of the entire effective action as in Eq. \eqref{sec3:action}
in what follows we consider solely the non-derivative and the two derivative 
part of the one loop correction given in Eq. \eqref{sec3:det_full}. 
Keeping in mind the remarks given in the introduction a scaling of couplings will be derived in the 
$MS$ scheme \cite{'tHooft:1973mm}. In this scheme the bare fields and the coupling constants
are related to the renormalized ones via the following formulae
\begin{subequations}
\begin{equation}
\begin{array}{c}
\label{sec3:renorm_constants}
\vf_B(\e) = \m^{1-\e/2}\vf(\m,\e) Z^{1/2}_\vf(g,\e) \ ,
\\[7pt]
{(\la_{2n})}_B(\e) =\m^{4-2n+(n-1)\e} g_{2n}(\m,\e) Z_{g_{2n}}(g,\e) \qquad \cbr{Z_{g_{2n}} \equiv Z_{\vf^{2n}}/Z^n_\vf}\ ,
\quad
n=1,2,\dots,\w \ ,
\end{array}
\end{equation}
and for gravitational coupling
\begin{equation}
\label{sec3:renorm_grav_lambda}
\ka^2_B(\e) = \mu^{\e-2}g_\ka(\m,\e) Z_{\ka}(g,\e)\ ,
\end{equation}
\end{subequations}
where in the above formula we have introduced the \emph{dimensionless} couplings $g_i$ and field. As we have
not computed a quantum corrections to the gravitational coupling its
renormalization constant is equal to one which entails a vanishing beta function for this coupling.
Remaining renormalization constants for couplings may be found from comparison of the simple pole terms in the
second line of Eq. \eqref{sec3:renorm_constants} and what we finally get is
$Z^{\scriptscriptstyle(1)}_{g_{2n}} = Z^{\scriptscriptstyle(1)}_{\vf^{2n}} - n Z^{\scriptscriptstyle(1)}_\vf \ $,
where the explicit forms of one loop parts of $Z^{\scriptscriptstyle(1)}_{\vf}$ and
$Z^{\scriptscriptstyle(1)}_{\vf^{2n}}$ are given in Eqs. (\ref{sec3:renorm_const_field}--\ref{sec3:renorm_const_vertices}).
Running of parameters $g_{2n}$ and anomalous dimension $\g_\vf(g)$ of the scalar
field may be found from the condition that the bare couplings
in Eqs. (\ref{sec3:renorm_constants} -- \ref{sec3:renorm_grav_lambda})
should not depend on $\m$ which, barring the running 
of gravitational coupling and taking the limit $\e\to0$,
amounts to the following formulae in $MS$ scheme
\begin{subequations}
\begin{equation}
\label{sec3:full_beta}
\b_{2n}(g) = \sbr{-(4-2n) + \g_{g_{2n}}(g) }g_{2n}\ , \quad n=1,2,\dots,\w \ ,
\end{equation}
where the second term in the above equation, to which we further refer as to 
anomalous dimensions for the scalar field couplings
$\g_{g_{2n}}$, and the anomalous dimension of a scalar field $\g_\vf$ take the general form
\begin{equation}
\label{sec3:anomalous_dims}
\g_\a (g) \equiv (1-\tfrac{3}{2}\d_{\a,\vf})
\sum_{j\in\{\ka,0,2n\}} a_j g_j \frac{\pt Z^{\scriptscriptstyle(1)}_{\a}(g)}{\pt g_j}\ ,
\quad \mbox{for}\quad \a = \vf, g_{2n}\ .
\end{equation}
\end{subequations}
In the above equations $a_j$ is a coefficient multiplying the DimReg
parameter $\e$ in an exponent of RG mass parameter $\mu$ in
Eqs. (\ref{sec3:renorm_constants}--\ref{sec3:renorm_grav_lambda}) and
$\d_{\a,\vf}$ is the Kronecker delta. By virtue of Eqs. (\ref{sec3:renorm_const_field}--\ref{sec3:renorm_const_vertices})
these formulae boil down to simple relations between corresponding anomalous
dimensions and coefficients of the simple poles of renormalization constants.
Hence the explicit form of the one loop anomalous dimensions for the scalar
field couplings from Eq. \eqref{sec3:full_beta} reads
\begin{subequations}
\begin{eqnarray}
\label{sec3:coupl_anomal_dim}
\nonr
\g^{(1L)}_{g_{2n}}(g)g_{2n} &=& \frac{1}{(4\pi)^2} \sum_{k=0}^n \binom{2n}{2k}
\Bigg\{
\frac{1}{2} g_{2(k+1)}g_{2(n-k+1)}-\cbr{C \frac{2(n-k)}{2k+1} + D}g_\ka g_{2(k+1)}g_{2(n-k)}
- B g^2_\ka g_{2k}g_{2(n-k)}
\Bigg\}
\\
&& {\hskip 50 pt}+\ 2n\frac{1}{(4\pi)^2}\cbr{E g_\ka g_2 + F g^2_\ka g_0} g_{2n}\ ,
\end{eqnarray}
and for the one loop anomalous dimension of the field one obtains
\begin{equation}
\label{sec3:field_anomal_dim}
\g^{(1L)}_\vf(g) =  \frac{1}{(4\pi)^2}\cbr{E g_\ka g_2  + F g^2_\ka g_0 }\ .
\end{equation}
\end{subequations}
The first term of the formula \eqref{sec3:coupl_anomal_dim}
is a pure nonlinear scalar field part of the one loop correction to the beta functions.
In the absence of gravitational interactions vanishing of the beta function
yields the FP. Apart from the mass parameter, all the scalar field couplings
obtain a positive contribution from quantum corrections and therefore the only
FP in this case is the one where all the couplings
vanish. This FP is a free field theory or Gaussian infrared
FP.\footnote{There are also other possible
fixed points apart from the Gaussian one that are parametrized by the mass parameter $g_2$ as
may be inferred from Eq. \eqref{sec3:coupl_anomal_dim} for $\b_2=0$ setting $g_0 = g_\ka =0$
and applying the solution to subsequent equations with vanishing beta functions. Although it provides
an infinite continuum number of FPs -- a fixed line -- the potentials have singularities
at some value of the field for all but zero mass parameters \cite{Morris:1996nx} and therefore
the only physically acceptable FP is the Gaussian FP \cite{Halpern:1996dh}.}
In order to asses whether this FP is stable or unstable with respect to the RG flow
one usually examines a flow of small perturbations about the FP determined
by means of linearized RG equations at this FP. However, in the $MS$ scheme
the lowest one loop order of anomalous dimension of coupling constant is quadratic in the
couplings and therefore at the Gaussian FP yields no information about its stability.

As for the gravitational contribution to beta functions let us first 
restrict ourselves to the polynomial potential containing
all up to quartic interaction. The results for different methods 
and gauges are summarized in the table \tabref{table2}.
\begin{table}
\begin{center}
\caption{Comparison of the one loop gravitational corrections to the beta
functions obtained in various methods (SEA $a=0$, VDEA $a=1$) and
gauges ($\xi ,b$). The notation is the following $\b_{2n} = \b^0_{2n} + \dl\b_{2n}$,
where $\b^0_{2n}=\b_{2n}(g_\ka =0, g_0 = 0)$ denotes the 
beta function for pure nonlinear scalar field theory,
whereas $\dl\b_{2n}$ represents the gravitational correction to it.}
\label{tab:table2}
\renewcommand{\arraystretch}{1.2}
\begin{tabular}{l@{\qquad}c@{\qquad}c}
\toprule
$(a,\xi,b)$ & $(4\pi)^2\dl\b^{\scriptscriptstyle(1L)}_2$ & $(4\pi)^2\dl\b^{\scriptscriptstyle(1L)}_4$ \\
\midrule
$(0,0,0)$ & $6\,{g}_{0}\,{g}_{2}\,{g}_{\kappa}^{2}$ & $-12\,{g}_{2}\,{g}_{4}\,{g}_{\kappa}
+\left( 6\,{g}_{0}\,{g}_{4}+18\,{g}_{2}^{2}\right)\,{g}_{\kappa}^{2}$ \\
$(0,1,0)$ & $8\,{g}_{0}\,{g}_{2}\,{g}_{\kappa}^{2}$ & $-8\,{g}_{2}\,{g}_{4}\,{g}_{\kappa}
+\left( 6\,{g}_{0}\,{g}_{4}+30\,{g}_{2}^{2}\right)\,{g}_{\kappa}^{2}$ \\
$(0,1,1)$ & $-2\,{g}_{2}^{2}\,{g}_{\kappa}+10\,{g}_{0}\,{g}_{2}\,{g}_{\kappa}^{2}$ & $-12\,{g}_{2}\,{g}_{4}\,{g}_{\kappa}
+\left( 10\,{g}_{0}\,{g}_{4}+30\,{g}_{2}^{2}\right) \,{g}_{\kappa}^{2}$ \\
$(0,0,1)$ & $-2\,{g}_{2}^{2}\,{g}_{\kappa}+4\,{g}_{0}\,{g}_{2}\,{g}_{\kappa}^{2}$ & $-16\,{g}_{2}\,{g}_{4}\,{g}_{\kappa}
+\left( 2\,{g}_{0}\,{g}_{4}+18\,{g}_{2}^{2}\right) \,{g}_{\kappa}^{2}$ \\
$(1,0,1)$ & $-\left({g}_{0}\,{g}_{4}+5\,{g}_{2}^{2}\right) \,{g}_{\kappa}+\tfrac{17}{2}\,{g}_{0}\,{g}_{2}\,{g}_{\kappa}^{2}$ &
$- 18\,{g}_{2}\,{g}_{4}\,{g}_{\kappa}
+\left( 10\,{g}_{0}\,{g}_{4}+21\,{g}_{2}^{2}\right) \,{g}_{\kappa}^{2}$ \\
\bottomrule
\end{tabular}
\end{center}
\end{table}
The last raw represents the unique gravitational
corrections to the beta functions. Recall that according to Eq. \eqref{sec3:origpot} 
and the rescaling $\la_0 = \m^4 g_0$ we have $g_0 g_\ka = g_\Lb$. Since 
a cosmological constant is an additional gravitational 
coupling we see that the leading gravitational corrections
enter the beta function for both mass and quartic
coupling with a negative sign. In the case of
positive cosmological constant the two contributions
give rise to a decrease of the effective couplings. On the
other hand the next to leading term which is of the form $\sim g_0 g^2_\ka = 2\Lb \ka^2$ 
produces the opposite effect. At low energy this term is negligible as compared to the 
leading contribution. At high energies, 
i.e. $g_\ka\sim (\m/M_P)^2 \sim 1$ it becomes important and competes with
the two negative contributions. In this case, however, prediction that hinges on the 
one loop beta function becomes unreliable, for higher order gravitational interactions
from the series defining the effective theory like $R^2$ must 
be taken into account. Hence, we conclude that the net effect of the
gravitational contribution in the adopted approximation 
gives rise to asymptotically free trend of running couplings.
On the other hand this contribution is small as compared to the pure scalar field one loop correction
which will dominate the running of scalar field effective couplings.
This remarks may be extended to the case of arbitrary
number of scalar field couplings. The only difference is that
now the beta function for the quartic coupling acquires a positive contribution from 
the non-renormalizable coupling $g_6$ in a pure scalar one loop correction 
and a negative contribution to the leading gravitational one
which may be found in appendix \ref{appB}.
Total one loop contribution to beta functions for non-renormalizable 
couplings is dominated by a canonical dimension term, and as such
governs the RG flow in vicinity of the Gaussian FP.

Before we proceed let us note that if we set $g_0 = 0$ then 
the second raw corresponds to the result found by Rodigast and
Schuster in Ref. \cite{Rodigast:2009zj}. Although their result has
been obtained by computing appropriate Feynman diagrams it is
tantamount to that obtained by means of the standard effective
action in harmonic gauge as we have done above. Taking into
account a different definition for gravitational constant ($\ka^2 =
\tilde{\ka}^2/2$ entails $g_\ka = {\tilde{g}}_\ka/2$) direct
comparison with Ref. \cite{Rodigast:2009zj} shows that the forms
of gravitational corrections coincide.

\subsection{The scalar field Gaussian fixed point}

The set of equations \eqref{sec3:full_beta} also admits a Gaussian FP.
Nevertheless it is interesting whether it admits a scalar field Gaussian FP (SGFP), with non-zero
gravitational couplings at the FP as well. Analysis of the unique form of RG equations
(see appendix \ref{appB}) reveals that, up to leading order in gravitational correction,
there is a FP solution where all but
$g_{0\ast}$ and $g_{2\ast}$ couplings vanish. However, it turns out to be unstable against
the addition of the next order gravitational correction. If we
include the next-to-leading gravitational correction the only non zero FP coupling appears to be $g_{0\ast}$.
Indeed, for if we put $g_2 = 0$ then vanishing of beta functions entails vanishing of all
the rest of scalar field couplings without the need of specifying the gravitational coupling.
This FP seems to be sought SGFP, although with the value $g_{0\ast}=8(4\pi)^2/7g^2_{\ka *}$
which is entirely out of reach the perturbation theory approach.
However, this FP may appear a spurious one as well, since if the next order corrections are 
added it might appear unstable. Nevertheless, it is likely that a genuine
FP for $g_0\neq 0$ does exists, for if we equate all the scalar
field couplings to zero, the only contribution will be that from gravitational coupling
which at each order, say $n$-th, will enter with a power of $(g_0g^2_\ka)^n$,
where $g_0g^2_\ka=2\ka^2\Lb$ is a dimensionless combination of gravitational couplings.
Given a flipping of the sign of gravitational coupling
with each order (as it happens at first and second order, see Eqs. \eqref{sec3:coupl_anomal_dim} )
it is conceivable that taking into account a complete series of
loop contributions we will eventually obtain an entire beta function with its zero in vicinity of the Gaussian
FP, that would then be a non-Gaussian FP for both a cosmological and a Newton coupling parameters
as the asymptotic safety scenario suggests.\footnote{The non-Gaussian FP was indeed found 
in the asymptotic safety scenario in Einstein-Hilbert truncation \cite{Reuter:1996cp,Donkin:2012ud}. } 
As it was mentioned earlier we have not calculated the beta function for
the gravitational coupling. Therefore it enters the RG equations as a small parameter.
Let us assume for both gravitational couplings to have a non zero values at the SGFP.
Such a situation takes place \textit{e.g.} in the asymptotic safety scenario \cite{Percacci:2003jz}.
Given a non-zero FP values for the gravitational couplings we are able to
examine the directions of the RG flow in vicinity of the SGFP.
Considering the unique form of the beta functions, when linearized
about the SGFP, yield the stability matrix that amounts to
\begin{equation}
\label{sec3:stability_matrix}
\frac{\pt\b_{2n}}{\pt g_{2m}}(g_{\ka *}, g_{0 *},0)
= \sbr{2n-4 + \tfrac{1}{(4\pi)^2}(7+\tfrac{3}{2}n)g^2_{\ka *} g_{0 *} }\d^{n}_{m}
- \tfrac{1}{(4\pi)^2}g_{\ka *} g_{0 *} \d^n_{m-1}\ , \quad n = 1,2,\dots, \w \ .
\end{equation}
Let us consider two cases: finite number of scalar field couplings $\w<\infty$ and 
infinite number of scalar field couplings $\w=\infty$.

\paragraph{The case of finite number of couplings $(\w<\infty)$.}
Assuming a finite number of scalar field vertex operators it is possible
to diagonalize the above stability matrix, eigenvalues of which are its diagonal elements.
Depending on the sign, these eigenvalues pinpoint a direction in which
an operator relative to a given eigenvalue flows in the course of the RG flow. These operators
that are attracted to the FP are termed relevant whereas those repelled from it -- irrelevant.
There are also a marginal operators that correspond a zero eigenvalue. As one may infer from 
the diagonal elements of Eq. \eqref{sec3:stability_matrix} for $g_{0 *}>0$ the gravitational correction
reduces number of relevant vertex operators.
In particular, a quartic operator being classically marginal,
due to gravitational correction \emph{becomes irrelevant}.
Thus the only relevant operator appears to be the mass operator.

\paragraph{The case of infinite number of couplings $(\w=\infty)$.}
As for the infinite number of couplings, it is possible to diagonalize the stability
matrix in Eq. \eqref{sec3:stability_matrix}. This time, however, off diagonal terms
also contribute the eigenvalue. It is worth mentioning that these terms derive
from the configuration space connection and are absent in standard background
field approach. The form of the stability matrix resembles that obtained in Wilson RG method in Refs.
\cite{Halpern:1994vw,Halpern:1995vf}. Therefore, making use of Eq. \eqref{sec3:stability_matrix},
it is possible to find a scalar field potential that has required properties of being physically non-trivial.
Solving the eigenvalue problem for small disturbances about the SGFP enables 
us to cast Eq. \eqref{sec3:stability_matrix} into a form
\begin{equation}
\label{sec3:recursion}
u_{2n+2} = \sbr{(4\pi)^2(2n-4)+(7+\tfrac{3}{2}n)g_{0\ast}g^2_{\ka\ast} - (4\pi)^2\theta}
u_{2n}/g_{0\ast}g_{\ka\ast}\ ,\quad n = 1,2,\dots\ ,
\end{equation}
where $u_i \equiv g_i - g_{i \ast}$ and $\theta$ is an eigenvalue. This is the recursion relation that starting
from $u_2$ allows one to express all the couplings in terms of $u_2$, $\theta$ and 
the fixed point values of gravitational couplings. Since at SGFP $g_{2n\ast}=0$ 
for $n>0$, this recursion relates all scalar field couplings to $g_2$. Explicitly,
\begin{subequations}
\begin{equation}
\label{sec3:recursion_result}
g_{2n} = \frac{\Gamma(a+n) \a^{n-1} }{\Gamma(a) a} g_2\ , \quad n>0
\end{equation}
where
\begin{equation}
\label{sec3:a_and_alpha}
a \equiv \frac{-4(4\pi)^2 + 7 g_{0\ast} g^2_{\ka\ast}-(4\pi)^2\theta}{2(4\pi)^2+\tfrac{3}{2}g_{0\ast} g^2_{\ka\ast}}\ ,
\qquad
\a \equiv \frac{4(4\pi)^2+3g_{0\ast} g^2_{\ka\ast}}{2g_{0\ast} g_{\ka\ast}}\ .
\end{equation}
\end{subequations}
The potential defined in Eq. \eqref{sec3:origpot} after making use of identity $(2n)! = 2^{2n} n! \Gamma(n+1/2)/\Gamma(1/2)$
may be rewritten as follows
\begin{equation}
\label{sec3:eigenpotential}
U_a(\vf) = g_0 +\tfrac{g_2}{\a a}\sbr{M\cbr{a,\tfrac{1}{2},\a\vf^2}-1}\ ,
\end{equation}
where $M(a,b,x)$ is the Kummer's function \cite{abramowitz+stegun}. Thus we have found a class 
of potentials, termed following Halpern and Huang \cite{Halpern:1994vw,Halpern:1995vf} \textit{eigenpotentials}.
Their shape is determined by the value of two parameters: $a$ and the mass parameter $g_2$.  
The latter, in turn is related to two gravitational
parameters $g_\ka$ and $g_0$, which is seen if we complete 
the stability matrix given in Eq. \eqref{sec3:stability_matrix}
with entries for $n=0$ that take the form
\begin{equation}
\frac{\pt \b_0}{\pt g_m}(g_{\ka\ast},g_{0\ast},0)
= -4 \d^0_m +\tfrac{1}{(4\pi)^2}\cbr{7 g_{0\ast} g^2_{\ka\ast}\d^0_m 
+ 7 g^2_{0\ast} g_{\ka\ast}\d^0_{m-\ka} - g_{0\ast} g_{\ka\ast}\d^0_{m-1} }\ .
\end{equation}
There is also a component $\pt \b_\ka/\pt g_m$. However, within the assumed approximation
in this paper this component is not known. The eigenvalue problem yields additional recursion relation
\begin{equation}
\label{sec3:recursion_g2}
g_2 = \a a u_0 + 7 g_{0\ast} u_\ka\ .
\end{equation}
This recursion relation will be modified if we include $\pt \b_\ka/\pt g_m$ which may be solved for $u_0$
in terms of $g_\ka$, $\theta$ and possibly $g_2$. In order to find a physically nontrivial potential the eigenvalue 
must be negative which implies for the two gravitational 
couplings to be attracted to their FP. Hence, a corresponding 
scalar field theory will be \emph{asymptotically free}.
Since the shape of the potential is determined by the 
parameter $a$ it is interesting to find its value
such that provides a possibility for symmetry breaking. Requirements for the potential to have 
this property are: $U'(0)<0$ and $U(\vf)>0$ for $\vf\gg 1$. For large $\vf$ the Kummer's function
behaves like $M(a,b,x)\sim \Gamma(b)x^{a-b} \ex^x/\Gamma(a)$. When applied to
Eq. \eqref{sec3:eigenpotential} these requirements entail the following conditions on $g_2$
\begin{equation*}
g_2 a<0 \quad \wedge \quad g_2/\Gamma(a)>0\ .
\end{equation*}
Since $g_2$ is related to gravitational couplings $u_0$ 
and $u_\ka$ as in Eq. \eqref{sec3:recursion_g2} there are many 
possibilities to fulfill these non-equalities. Let us consider 
one of them and assume for simplicity that $g_2>0$. This implies that
$a<0$ and according to properties of the Gamma function we get $a\in (-2 k,-2k+1)$ for $k>0$. 
From Eq. \eqref{sec3:a_and_alpha} for $\theta<0$ one may infer that $a$ must 
fall at most into the interval $a\in(-2,-1)$. If we take the FP value for $g_0$ obtained 
from vanishing of the beta function for $n=0$ in Eq. \eqref{sec3:full_beta} 
which amounts $g_{0\ast}=8(4\pi)^2/7g^2_{\ka\ast}$ 
then we obtain $a=7(4-\theta)/28$, a value that falls outside the mentioned 
interval. However, this value for $a$ derives from the one loop approximation
to the beta function. Nevertheless, it is conceivable that
for the full beta function $a<0$ and therefore belongs to this interval.

\section{Summary and conclusions \label{sec5:title}}

In this paper we reconsidered quantum gravitational corrections to renormalization of
the scalar field couplings, and the effect they have on their running that had 
been touched upon earlier in different contexts by many authors
\cite{Barvinsky:1993zg,Griguolo:1995db,Percacci:2003jz,
Narain:2009fy,Narain:2009gb,He:2010mt,Rodigast:2009zj,
Mackay:2009cf,Anber:2010uj,Steinwachs:2011zs}. The reason we undertook this 
task was to investigate whether the influence of quantum 
gravitational fluctuations is capable to resolve the problem
of triviality in an interacting quantum scalar field theory.
We searched for these corrections within the effective field 
theory approach to quantum gravity and confined ourselves to a cosmological
constant and Ricci scalar. A scalar field potential is 
assumed to have a $\mathbb{Z}_2$ symmetric and analytic form.
As we performed computations in the flat background metric 
all the operators with nonminimally coupled scalar fields to the gravity were discarded. 
This subject was recently discussed within the four dimensional 
massive scalar field theory with quartic interaction by means of the 
off the mass shell Feynman diagram computations in Ref. \cite{Rodigast:2009zj}.
A sign of the beta function determine the direction 
of a change an effective coupling undergoes with energy.
It may, however, vary depending on the chosen gauge which usually takes
place in off the mass shell computations. In order to enable an adequate 
treatment of the diffeomorphism symmetry of gravitational field
as well as to obtain a unique result in the off shell computations
we used a geometric formulation of the method of background field, namely the Vilkovisky-DeWitt 
effective action. Using this method 
we derived the unique, viz. gauge independent beta functions for all dimensionless coupling 
parameters of the theory defined in the $MS$ scheme. Since we restricted our considerations to the flat 
background, the beta function for the Newton coupling parameter $g_\ka\propto \ka^2$ assumed zero value. 
The analysis of the system of RG equations for the scalar field couplings revealed that the leading 
order gravitational correction to all the beta functions of 
scalar field couplings act in the direction of asymptotic 
freedom as found in Ref. \cite{Rodigast:2009zj} in harmonic gauge,
although in a different form. In addition to the contribution 
from the Newton constant there is the one coming from the cosmological
constant. In the case of quartic coupling which is the marginal coupling the presence of cosmological constant
modifies the asymptotically free trend which is due to the leading gravitational contribution. 
A positive cosmological constant enhances the effect of the leading gravitational correction.
However, this effect is small as compared to pure scalar field contribution.
As for the rest of the scalar field couplings a dominating contribution to beta functions
comes from their canonical dimensions. Thus their running does not change much in the 
presence of gravitational interactions. 

Moreover, we also found that RG equations admit 
another FP with non zero FP values solely for
both gravitational couplings $\Lb$ and $\ka^2$ that is the 
scalar field Gaussian FP (SGFP). Since we did not determine 
the form of the beta function for $\ka^2$ this coupling entered the computations as a free parameter.
In order to examine what consequences it may have we assumed it to take a nonzero value at the FP.
This, in view of found RG equations, entails a nonzero FP value for $\Lb$. 
We found through examination of stability matrix at SGFP
that for a finite number of scalar field vertex operators gravitational corrections 
render them more irrelevant. Specifically, a quartic 
operator being marginal in the absence of gravitational interactions is made irrelevant
due to gravitational contribution. These conclusions were also 
met in Ref. \cite{Percacci:2003jz} where the theory of scalar field non-minimally coupled to gravity
was explored within the effective average action.

We also considered the case of infinite many 
scalar field interactions examined earlier in a pure interacting scalar field theory 
by Halpern and Huang in Refs. \cite{Halpern:1994vw,Halpern:1995vf}.
The reason for this was to explore a possible
nontrivial directions with respect to the RG flow in the space of all 
scalar field coupling parameters defined in $MS$ scheme in the presence of gravitational interactions. 
In order to do this we looked for the solution of linearized
RG equations for small disturbances about the SGFP. The stability matrix 
found in this way is bidiagonal. The second diagonal comes from the Vilkovisky-DeWitt configuration 
space connection and is absent in the stability matrix derived within standard 
formulation of the background field method. 
Owing to the bidiagonal form the eigenvalue 
problem boiled down to the recursion relation for all the couplings.
As a result we found a class of potentials termed eigenpotentials parametrized by the eigenvalue 
and that depend merely on the two gravitational couplings.
In order for the scalar field theory to be nontrivial the eigenvalue must be negative.
Hence the theory with the eigenpotential corresponding to this eigenvalue is asymptotically free.
The shape of the eigenpotentials is entirely determined by some parameter $a$ which is 
a linear function of the eigenvalue and nonlinear function 
of FP values of both gravitational coupling parameters $g_\ka$ and $g_0$. 
The most appealing eigenpotentials are those that admit the symmetry breaking.
This substantially constrains the set of possible values for the shape parameter $a$. 
In the case considered in this paper it is confined to a certain open intervals 
of the negative part of $\mathbb{R}$. Taking the FP value of $g_0$ found in this one loop 
approximation to $\b_0$ the shape parameter is positive. If taken at face value this would imply 
that the theory with nontrivial eigenpotentials does not admit the symmetry breaking shapes.
However, this may not be the case if we take the FP value of $g_0$ obtained from the full 
beta function. Thus we found a class of 
scalar field potentials -- gravitationally modified Halpern-Huang potentials -- 
that are non-polynomial and that have features making an interacting scalar field theory nontrivial
provided that there exists a non-zero fixed point value for the two gravitational 
couplings, namely the Newton constant and the cosmological constant.
A non-perturbative studies of Einstein quantum gravity \cite{Reuter:1996cp,Donkin:2012ud} indicate 
that a non-zero FP values for the two gravitational couplings may indeed exist.
Interestingly, this result was derived within the $MS$ scheme.
Nevertheless, it has a universal validity, as the FP's as well as eigenvalues
do not depend on a specific definition of coupling constants.
Since this result hinges on a continuum rather then quantized 
eigenvalue as well as non-polynomial potential the remarks and the caveats mentioned in the 
first paragraph of section \ref{sec1} also apply in this case.

The analysis performed in this paper does not allow for 
operators with scalar field non-minimally coupled to gravity, 
which is acceptable in adopted approximation, 
\textit{i.e.} flat background metric. However, 
in curved spacetime non-minimal coupling to gravity is
required for reason of renormalizability. From this 
point of view, investigations just performed
are pertaining to the subspace of the full coupling 
parameter space. It is therefore interesting to examine
how the presence of non-minimal couplings affect 
the triviality issue when considered in the framework
of Vilkovisky-DeWitt effective action. Specifically, 
whether in case of infinite many scalar field couplings
it is possible to find potentials with nontrivial properties at high energies. 
This task will be undertaken in a separate paper \cite{Pietrykowski}.

\subsection*{Acknowledgements}

I am deeply indebted to Professor Z.T. Haba from Wroclaw University, 
Professor E.A. Ivanov from BLTP, JINR,
and Dr. M.R. Piatek from Szczecin University for careful reading 
of the manuscript and for many valuable comments and critical remarks.
I am grateful for kind hospitality and interesting 
conversation to Dr. F. S. Nogueira from Free University of Berlin
where the early stage of this work was presented during the lecture.
I am also thankful to Professor J.M. Pawlowski from Heidelberg University
for enlightening discussion on some aspects of the Vilkovisky-DeWitt formalism.

\newpage

\appendix

\section{The definitions and notation \label{appA}}

\setcounter{equation}{0}
\renewcommand{\theequation}{A.\arabic{equation}}

Evaluation of momentum integrals with explicit indices in integrated components of momenta
results in multiindex generalized deltas defined below. 
The meaning of $\unit{n}$ employed in section \ref{sec3:title} is the following
\begin{equation}
\label{appA:jedynka}
\begin{array}{c}
{(\unit{1})}_{\a\b} \equiv \d_{\a\b}\ ,
\quad {(\unit{2})}_{\a\b,\m\n} \equiv \d_{\a\b,\m\n}
\equiv  \d_{\a(\m}\d_{\n)\b} = \tfrac{1}{2}\cbr{\d_{\a\m}\d_{\b\n}+\d_{\a\n}\d_{\b\m}}
\\[10pt]
{(\unit{3})}_{\a\b,\m\n,\r\s} \equiv \d_{\a\b,\m\n,\r\s} \equiv \d_{\a\b,(\m(\r}\d_{\s)\n)}\ ,
\\[10pt]
{(\unit{n})}_{\a_1\b_1,\a_2\b_2,\ldots,\a_{n-1}\b_{n-1},\a_n\b_n}
\equiv \d_{\a_1\b_1,\a_2\b_2,\ldots,\a_{n-1}\b_{n-1},\a_n\b_n}\ ,
\end{array}
\end{equation}
where the indices embraced with curl brackets indicate that these indices are to be symmetrized, that is
\begin{equation}
\label{appA:symmetrization}
A_{(\a\b)}\equiv \tfrac{1}{2}\cbr{A_{\a\b}+A_{\b\a}}\ .
\end{equation}
In particular the symbol we have used to compute the last ingredient 
of the trace preceding Eq. \eqref{sec3:resultQQQQ} takes the form
\begin{equation}
\label{appA:deltabig}
\d_{\a\b,\g\la,\m\n,\r\s} = \d_{(\s|(\a}\d_{\b)(\g}\d_{\la)(\m}\d_{\n)|\r)}
+ \d_{(\n|(\a}\d_{\b)(\g}\d_{\la)(\r}\d_{\s)|\m)} + \d_{(\s|(\a}\d_{\b)(\m}\d_{\n)(\g}\d_{\la)|\r)},
\end{equation}
where indices between the bars are excluded from symmetrization procedure.
The representation of the last formula in the above expression in terms of Kronecker delta is highly nontrivial
and will not be given here. For the sake of brevity we introduce doubled index $i_l \equiv (\m_l\n_l)$.
The above defined quantities satisfy the following identities.
\bse
\ba
\label{id1}
(\unit{n})_{i_1,\ldots,i_m}\equiv \d_{i_1,\ldots ,i_{k-1},i_k,i_{k+1},\ldots,i_n}
&=&\d_{i_k,j}\d^{j}_{i_1,\ldots,i_{k-1},i_{k+1},\ldots,i_n},
\quad \d_{(i_1,\ldots ,i_n)}=\d_{i_1,\ldots ,i_n},
\\
\label{id2}
\d^{i_k}\d_{i_1,\ldots ,i_{k-1},i_k,i_{k+1},\ldots,i_n}
&=&\d_{i_1,\ldots ,i_{k-1},i_{k+1},\ldots,i_n}.
\ea
\ese
DeWitt configuration space metric defined in Eq. \eqref{sec3:dewitt_metric} 
and its inverse can be written in the flat $n -$dimensional Euclidian spacetime as
\be
\gr{i}{j} = \tfrac{1}{4}\cbr{2\d^{i,j}-\d^i\d^j},\quad \igr{i}{j} = 2\d_{i,j}-\tfrac{2}{n-2}\d_i\d_j.
\ee

\section{The unique $\beta$ functions for scalar--gravity system \label{appB}}

In this appendix we present explicit form of the beta functions
obtained in the Eq. \eqref{sec3:full_beta} and \eqref{sec3:coupl_anomal_dim}
for unique values of coefficients given in \tabref{table1}, i.e. VDEA
for $n=5$. The full beta function can be split into two parts, the one for pure scalar 
field theory and that coming from gravitational corrections, namely
$$
\b_{2n}(g) = \b^0_{2n}(g) + \dl\b_{2n}(g)\ ,
$$
where
\begin{align*}
\b^0_0 &=-4 g_0+\tfrac{1}{32\pi^2}g^2_2 \ ,
\\[5pt]
{\b}^0_{2}&=-2\,{g}_{2}+\tfrac{1}{16\,{\pi}^{2}}{g}_{2}\,{g}_{4}\ ,
\\[5pt]
{\b}^0_{4}&=\tfrac{1}{16\,{\pi}^{2}}({g}_{2}\,{g}_{6}+3\,{g}_{4}^{2}) \ ,
\\[5pt]
{\b}^0_{6}&=2\,{g}_{6}+\tfrac{1}{16\,{\pi}^{2}}({g}_{2}\,{g}_{8}+15\,{g}_{4}\,{g}_{6})\ ,
\\[5pt]
{\b}^0_{8}&=4\,{g}_{8}+\tfrac{1}{16\,{\pi}^{2}}({g}_{2}\,{g}_{10}+28\,{g}_{4}\,{g}_{8}+35\,{g} _ {6}^{2})\ ,
\\[5pt]
{\b}^0_{10}&=6\,{g}_{10}+\tfrac{1}{16\,{\pi}^{2}}({g}_{2}\,{g}_{12}+45\,{g}_{4}\,{g}_{10}+210\,{g}_{6}\,{g}_{8})\ ,
\\
&\vdots
\end{align*}
and
\begin{align*}
\dl \b_0(g)&=\tfrac{1}{16\pi^2}\sbr{-g_0 g_2 g_\ka + \tfrac{7}{2}g^2_0 g^2_\ka} \ ,
\\[5pt]
\dl \b_2(g) &= \tfrac{1}{16\,{\pi}^{2}}
\sbr{-\left({g}_{0}\,{g}_{4}+5{g}_{2}^{2}\right){g}_{\kappa}
+\tfrac{17}{2}{g}_{0}\,{g}_{2}\,{g}_{\kappa}^{2}} \ ,
\\[5pt]
\dl \b_4(g) &= \tfrac{1}{16\,{\pi}^{2}}
\sbr{-\left({g}_{0}\,{g}_{6}+18\,{g}_{2}\,{g}_{4}\right){g}_{\kappa}
+\left( 10\,{g}_{0}\,{g}_{4}+21\,{g}_{2}^{2}\right)
\,{g}_{\kappa}^{2}}\ ,
\\[5pt]
\dl \b_6(g) &= \tfrac{1}{16\,{\pi}^{2}}
\sbr{-\left({g}_{0}\,{g}_{8}+\tfrac{65}{2}\,{g}_{2}\,{g}_{6}+30\,{g}_{4}^{2}\right){g}_{\kappa}
+\left(\tfrac{23}{2}{g}_{0}{g}_{6}+105\,{g}_{2}\,{g}_{4}\right){g}_{\kappa}^{2}}\ ,
\\[5pt]
\dl\b_{8}(g) &= \tfrac{1}{16\,{\pi}^{2}}
\sbr{-\left({g}_{0}\,{g}_{10}+51\,{g}_{2}\,{g}_{8}+182\,{g}_{4}\,{g}_{6}\right)
\,{g}_{\kappa}+\left(13\,{g}_{0}\,{g}_{8}+196\,{g}_{2}\,{g}_{6}+245\,{g}_{4}^{2}\right)
\,{g}_{\kappa}^{2}}\ ,
\\[5pt]
\dl\b_{10}(g) &= \tfrac{1}{16\,{\pi}^{2}}
\Big[-\left(\,{g}_{0}\,{g}_{12}+\tfrac{147}{2}\,{g}_{2}\,{g}_{10}+435\,{g}_{4}\,{g}_{8}+399\,{g}_{6}^{2}\right){g}_{\kappa}
\\
&{\hskip 10pt}+\left(\tfrac{29}{2}\,{g}_{0}\,{g}_{10}+315\,{g}_{2}\,{g}_{8}+1470\,{g}_{4}\,{g}_{6}\right){g}_{\kappa}^{2}\Big]\ ,
\\
&\vdots
\end{align*}

\providecommand{\href}[2]{#2}\begingroup\raggedright\endgroup


\begin{thebibliography}{10}

\bibitem{Dolan:1973qd}
L.~Dolan and R.~Jackiw, {\it {Symmetry Behavior at Finite Temperature}},  {\em
  Phys.Rev.} {\bf D9} (1974) 3320--3341.

\bibitem{Coleman:1974jh}
S.~R. Coleman, R.~Jackiw, and H.~D. Politzer, {\it {Spontaneous Symmetry
  Breaking in the O(N) Model for Large N*}},  {\em Phys.Rev.} {\bf D10} (1974)
  2491.

\bibitem{Wilson:1971bg}
K.~G. Wilson, {\it {Renormalization group and critical phenomena. 1.
  Renormalization group and the Kadanoff scaling picture}},  {\em Phys.Rev.}
  {\bf B4} (1971) 3174--3183.

\bibitem{Wilson:1973jj}
K.~Wilson and J.~B. Kogut, {\it {The Renormalization group and the epsilon
  expansion}},  {\em Phys.Rept.} {\bf 12} (1974) 75--200.

\bibitem{Hasenfratz:1985dm}
A.~Hasenfratz and P.~Hasenfratz, {\it {Renormalization Group Study of Scalar
  Field Theories}},  {\em Nucl.Phys.} {\bf B270} (1986) 687--701.

\bibitem{Callaway:1988ya}
D.~J. Callaway, {\it {Triviality Pursuit: Can Elementary Scalar Particles
  Exist?}},  {\em Phys.Rept.} {\bf 167} (1988) 241.

\bibitem{Halpern:1994vw}
K.~Halpern and K.~Huang, {\it {Fixed point structure of scalar fields}},  {\em
  Phys.Rev.Lett.} {\bf 74} (1995) 3526--3529,
  [\href{http://xxx.lanl.gov/abs/hep-th/9406199}{{\tt hep-th/9406199}}].

\bibitem{Halpern:1995vf}
K.~Halpern and K.~Huang, {\it {Nontrivial directions for scalar fields}},  {\em
  Phys.Rev.} {\bf D53} (1996) 3252--3259,
  [\href{http://xxx.lanl.gov/abs/hep-th/9510240}{{\tt hep-th/9510240}}].

\bibitem{Morris:1996nx}
T.~R. Morris, {\it {On the fixed point structure of scalar fields}},  {\em
  Phys.Rev.Lett.} {\bf 77} (1996) 1658,
  [\href{http://xxx.lanl.gov/abs/hep-th/9601128}{{\tt hep-th/9601128}}].

\bibitem{Halpern:1996dh}
K.~Halpern and K.~Huang, {\it {Reply to: Comment on 'Fixed point structure of
  scalar fields'}},  {\em Phys.Rev.Lett.} {\bf 77} (1996) 1659.

\bibitem{Halpern:1997gn}
K.~Halpern, {\it {Cross-section and effective potential in asymptotically free
  scalar field theories}},  {\em Phys.Rev.} {\bf D57} (1998) 6337--6341,
  [\href{http://xxx.lanl.gov/abs/hep-th/9708124}{{\tt hep-th/9708124}}].

\bibitem{Gies:2000xr}
H.~Gies, {\it {Flow equation for Halpern-Huang directions of scalar O(N)
  models}},  {\em Phys.Rev.} {\bf D63} (2001) 065011,
  [\href{http://xxx.lanl.gov/abs/hep-th/0009041}{{\tt hep-th/0009041}}].

\bibitem{Branchina:2000jp}
V.~Branchina, {\it {Nonperturbative renormalization group potentials and
  quintessence}},  {\em Phys.Rev.} {\bf D64} (2001) 043513,
  [\href{http://xxx.lanl.gov/abs/hep-ph/0002013}{{\tt hep-ph/0002013}}].

\bibitem{Huang:2011xg}
K.~Huang, H.-B. Low, and R.-S. Tung, {\it {Scalar Field Cosmology I: Asymptotic
  Freedom and the Initial-Value Problem}},  {\em Class.Quant.Grav.} {\bf 29}
  (2012) 155014, [\href{http://xxx.lanl.gov/abs/1106.5282}{{\tt
  arXiv:1106.5282}}].

\bibitem{Huang:2011xha}
K.~Huang, H.-B. Low, and R.-S. Tung, {\it {Scalar Field Cosmology II:
  Superfluidity and Quantum Turbulence}},
  \href{http://xxx.lanl.gov/abs/1106.5283}{{\tt arXiv:1106.5283}}.

\bibitem{Rosten:2008ts}
O.~J. Rosten, {\it {Triviality from the Exact Renormalization Group}},  {\em
  JHEP} {\bf 0907} (2009) 019, [\href{http://xxx.lanl.gov/abs/0808.0082}{{\tt
  arXiv:0808.0082}}].

\bibitem{Cheng:1973nv}
T.~Cheng, E.~Eichten, and L.-F. Li, {\it {Higgs Phenomena in Asymptotically
  Free Gauge Theories}},  {\em Phys.Rev.} {\bf D9} (1974) 2259.

\bibitem{'tHooft:1974bx}
G.~'t~Hooft and M.~Veltman, {\it {One loop divergencies in the theory of
  gravitation}},  {\em Annales Poincare Phys.Theor.} {\bf A20} (1974) 69--94.

\bibitem{Goroff:1985th}
M.~H. Goroff and A.~Sagnotti, {\it {The Ultraviolet Behavior of Einstein
  Gravity}},  {\em Nucl.Phys.} {\bf B266} (1986) 709.

\bibitem{Donoghue:1994dn}
J.~F. Donoghue, {\it {General relativity as an effective field theory: The
  leading quantum corrections}},  {\em Phys.Rev.} {\bf D50} (1994) 3874--3888,
  [\href{http://xxx.lanl.gov/abs/gr-qc/9405057}{{\tt gr-qc/9405057}}].

\bibitem{Weinberg1979}
S.~Weinberg, {\it Ultraviolet divergencies in quantum theories of gravitation},
   in {\em General Relativity: An Einstein Centenary Survey} (S.~Hawking and
  W.~Israel, eds.).
\newblock Cambridge University Press, Cambridge, U.K.; New York, U.S.A., 1979.

\bibitem{Georgi:1994qn}
H.~Georgi, {\it {Effective field theory}},  {\em Ann.Rev.Nucl.Part.Sci.} {\bf
  43} (1993) 209--252.

\bibitem{Pich:1998xt}
A.~Pich, {\it {Effective field theory: Course}},
  \href{http://xxx.lanl.gov/abs/hep-ph/9806303}{{\tt hep-ph/9806303}}.

\bibitem{Donoghue:1995cz}
J.~F. Donoghue, {\it {Introduction to the effective field theory description of
  gravity}},  \href{http://xxx.lanl.gov/abs/gr-qc/9512024}{{\tt
  gr-qc/9512024}}.

\bibitem{Burgess:2003jk}
C.~Burgess, {\it {Quantum gravity in everyday life: General relativity as an
  effective field theory}},  {\em Living Rev.Rel.} {\bf 7} (2004) 5,
  [\href{http://xxx.lanl.gov/abs/gr-qc/0311082}{{\tt gr-qc/0311082}}].

\bibitem{Donoghue:2012zc}
J.~F. Donoghue, {\it {The effective field theory treatment of quantum
  gravity}},  \href{http://xxx.lanl.gov/abs/1209.3511}{{\tt arXiv:1209.3511}}.

\bibitem{Robinson:2005fj}
S.~P. Robinson and F.~Wilczek, {\it {Gravitational correction to running of
  gauge couplings}},  {\em Phys.Rev.Lett.} {\bf 96} (2006) 231601,
  [\href{http://xxx.lanl.gov/abs/hep-th/0509050}{{\tt hep-th/0509050}}].

\bibitem{Pietrykowski:2006xy}
A.~R. Pietrykowski, {\it {Gauge dependence of gravitational correction to
  running of gauge couplings}},  {\em Phys.Rev.Lett.} {\bf 98} (2007) 061801,
  [\href{http://xxx.lanl.gov/abs/hep-th/0606208}{{\tt hep-th/0606208}}].

\bibitem{Ebert:2007gf}
D.~Ebert, J.~Plefka, and A.~Rodigast, {\it {Absence of gravitational
  contributions to the running Yang-Mills coupling}},  {\em Phys.Lett.} {\bf
  B660} (2008) 579--582, [\href{http://xxx.lanl.gov/abs/0710.1002}{{\tt
  arXiv:0710.1002}}].

\bibitem{Toms:2007sk}
D.~J. Toms, {\it {Quantum gravity and charge renormalization}},  {\em
  Phys.Rev.} {\bf D76} (2007) 045015,
  [\href{http://xxx.lanl.gov/abs/0708.2990}{{\tt arXiv:0708.2990}}].

\bibitem{Felipe:2011rs}
J.~Felipe, L.~Brito, M.~Sampaio, and M.~Nemes, {\it {Quantum gravitational
  contributions to the beta function of quantum electrodynamics}},  {\em
  Phys.Lett.} {\bf B700} (2011) 86--89,
  [\href{http://xxx.lanl.gov/abs/1103.5824}{{\tt arXiv:1103.5824}}].

\bibitem{Toms:2008dq}
D.~J. Toms, {\it {Cosmological constant and quantum gravitational corrections
  to the running fine structure constant}},  {\em Phys.Rev.Lett.} {\bf 101}
  (2008) 131301, [\href{http://xxx.lanl.gov/abs/0809.3897}{{\tt
  arXiv:0809.3897}}].

\bibitem{Toms:2009vd}
D.~J. Toms, {\it {Quantum gravity, gauge coupling constants, and the
  cosmological constant}},  {\em Phys.Rev.} {\bf D80} (2009) 064040,
  [\href{http://xxx.lanl.gov/abs/0908.3100}{{\tt arXiv:0908.3100}}].

\bibitem{Toms:2010vy}
D.~J. Toms, {\it {Quantum gravitational contributions to quantum
  electrodynamics}},  {\em Nature} {\bf 468} (2010) 56--59,
  [\href{http://xxx.lanl.gov/abs/1010.0793}{{\tt arXiv:1010.0793}}].

\bibitem{He:2010mt}
H.-J. He, X.-F. Wang, and Z.-Z. Xianyu, {\it {Gauge-Invariant Quantum Gravity
  Corrections to Gauge Couplings via Vilkovisky-DeWitt Method and Gravity
  Assisted Gauge Unification}},  {\em Phys.Rev.} {\bf D83} (2011) 125014,
  [\href{http://xxx.lanl.gov/abs/1008.1839}{{\tt arXiv:1008.1839}}].

\bibitem{Tang:2011gz}
Y.~Tang and Y.-L. Wu, {\it {Gravitational Contributions to Gauge Green's
  Functions and Asymptotic Free Power-Law Running of Gauge Coupling}},  {\em
  JHEP} {\bf 1111} (2011) 073, [\href{http://xxx.lanl.gov/abs/1109.4001}{{\tt
  arXiv:1109.4001}}].

\bibitem{Tang:2010cr}
Y.~Tang and Y.-L. Wu, {\it {Quantum Gravitational Contributions to Gauge Field
  Theories}},  {\em Commun.Theor.Phys.} {\bf 57} (2012) 629--636,
  [\href{http://xxx.lanl.gov/abs/1012.0626}{{\tt arXiv:1012.0626}}].

\bibitem{Ellis:2010rw}
J.~Ellis and N.~E. Mavromatos, {\it {On the Interpretation of Gravitational
  Corrections to Gauge Couplings}},  {\em Phys.Lett.} {\bf B711} (2012)
  139--142, [\href{http://xxx.lanl.gov/abs/1012.4353}{{\tt arXiv:1012.4353}}].

\bibitem{Anber:2010uj}
M.~M. Anber, J.~F. Donoghue, and M.~El-Houssieny, {\it {Running couplings and
  operator mixing in the gravitational corrections to coupling constants}},
  {\em Phys.Rev.} {\bf D83} (2011) 124003,
  [\href{http://xxx.lanl.gov/abs/1011.3229}{{\tt arXiv:1011.3229}}].

\bibitem{Anber:2011ut}
M.~M. Anber and J.~F. Donoghue, {\it {On the running of the gravitational
  constant}},  {\em Phys.Rev.} {\bf D85} (2012) 104016,
  [\href{http://xxx.lanl.gov/abs/1111.2875}{{\tt arXiv:1111.2875}}].

\bibitem{Toms:2011zza}
D.~J. Toms, {\it {Quadratic divergences and quantum gravitational contributions
  to gauge coupling constants}},  {\em Phys.Rev.} {\bf D84} (2011) 084016.

\bibitem{NK2012861}
N.~Nielsen, {\it {The Maxwell-Einstein system, Ward identities and the
  Vilkovisky construction}},  {\em Annals of Physics} {\bf 327} (2012), no.~3
  861 -- 892.

\bibitem{Daum:2009dn}
J.-E. Daum, U.~Harst, and M.~Reuter, {\it {Running Gauge Coupling in
  Asymptotically Safe Quantum Gravity}},  {\em JHEP} {\bf 1001} (2010) 084,
  [\href{http://xxx.lanl.gov/abs/0910.4938}{{\tt arXiv:0910.4938}}].

\bibitem{Folkerts:2011jz}
S.~Folkerts, D.~F. Litim, and J.~M. Pawlowski, {\it {Asymptotic freedom of
  Yang-Mills theory with gravity}},  {\em Phys.Lett.} {\bf B709} (2012)
  234--241, [\href{http://xxx.lanl.gov/abs/1101.5552}{{\tt arXiv:1101.5552}}].

\bibitem{Kazakov:1987jp}
D.~Kazakov, {\it {On a generalization of renormalization group equations to
  quantum field theories of an arbitrary type}},  {\em Theor.Math.Phys.} {\bf
  75} (1988) 440--442.

\bibitem{Barvinsky:1993zg}
A.~Barvinsky, A.O.~Kamenshchik and I.~Karmazin, {\it {The Renormalization group
  for nonrenormalizable theories: Einstein gravity with a scalar field}},  {\em
  Phys.Rev.} {\bf D48} (1993) 3677--3694,
  [\href{http://xxx.lanl.gov/abs/gr-qc/9302007}{{\tt gr-qc/9302007}}].

\bibitem{Steinwachs:2011zs}
C.~F. Steinwachs and A.~Y. Kamenshchik, {\it {One-loop divergences for gravity
  non-minimally coupled to a multiplet of scalar fields: calculation in the
  Jordan frame. I. The main results}},  {\em Phys.Rev.} {\bf D84} (2011)
  024026, [\href{http://xxx.lanl.gov/abs/1101.5047}{{\tt arXiv:1101.5047}}].

\bibitem{Griguolo:1995db}
L.~Griguolo and R.~Percacci, {\it {The Beta functions of a scalar theory
  coupled to gravity}},  {\em Phys.Rev.} {\bf D52} (1995) 5787--5798,
  [\href{http://xxx.lanl.gov/abs/hep-th/9504092}{{\tt hep-th/9504092}}].

\bibitem{Percacci:2003jz}
R.~Percacci and D.~Perini, {\it {Asymptotic safety of gravity coupled to
  matter}},  {\em Phys.Rev.} {\bf D68} (2003) 044018,
  [\href{http://xxx.lanl.gov/abs/hep-th/0304222}{{\tt hep-th/0304222}}].

\bibitem{Reuter:1996cp}
M.~Reuter, {\it {Nonperturbative evolution equation for quantum gravity}},
  {\em Phys.Rev.} {\bf D57} (1998) 971--985,
  [\href{http://xxx.lanl.gov/abs/hep-th/9605030}{{\tt hep-th/9605030}}].

\bibitem{Narain:2009fy}
G.~Narain and R.~Percacci, {\it {Renormalization Group Flow in Scalar-Tensor
  Theories. I}},  {\em Class.Quant.Grav.} {\bf 27} (2010) 075001,
  [\href{http://xxx.lanl.gov/abs/0911.0386}{{\tt arXiv:0911.0386}}].

\bibitem{Falkenberg:1996bq}
S.~Falkenberg and S.~D. Odintsov, {\it {Gauge dependence of the effective
  average action in Einstein gravity}},  {\em Int.J.Mod.Phys.} {\bf A13} (1998)
  607--623, [\href{http://xxx.lanl.gov/abs/hep-th/9612019}{{\tt
  hep-th/9612019}}].

\bibitem{Souma:2000vs}
W.~Souma, {\it {Gauge and cutoff function dependence of the ultraviolet fixed
  point in quantum gravity}},
  \href{http://xxx.lanl.gov/abs/gr-qc/0006008}{{\tt gr-qc/0006008}}.

\bibitem{Rodigast:2009zj}
A.~Rodigast and T.~Schuster, {\it {Gravitational Corrections to Yukawa and
  $\phi^4$ Interactions}},  {\em Phys.Rev.Lett.} {\bf 104} (2010) 081301,
  [\href{http://xxx.lanl.gov/abs/0908.2422}{{\tt arXiv:0908.2422}}].

\bibitem{Mackay:2009cf}
P.~T. Mackay and D.~J. Toms, {\it {Quantum gravity and scalar fields}},  {\em
  Phys.Lett.} {\bf B684} (2010) 251--255,
  [\href{http://xxx.lanl.gov/abs/0910.1703}{{\tt arXiv:0910.1703}}].

\bibitem{Chang:2012zzo}
H.-R. Chang, W.-T. Hou, and Y.~Sun, {\it {Gravitational corrections to $\phi^4$
  theory with spontaneously broken symmetry}},  {\em Phys.Rev.} {\bf D85}
  (2012) 124025, [\href{http://xxx.lanl.gov/abs/1207.5981}{{\tt
  arXiv:1207.5981}}].

\bibitem{Vilkovisky:1984st}
G.~Vilkovisky, {\it {The Unique Effective Action in Quantum Field Theory}},
  {\em Nucl.Phys.} {\bf B234} (1984) 125--137.

\bibitem{'tHooft:1973mm}
G.~'t~Hooft, {\it {Dimensional regularization and the renormalization group}},
  {\em Nucl.Phys.} {\bf B61} (1973) 455--468.

\bibitem{DeWitt:1967ub}
B.~S. DeWitt, {\it {Quantum Theory of Gravity. 2. The Manifestly Covariant
  Theory}},  {\em Phys.Rev.} {\bf 162} (1967) 1195--1239.

\bibitem{Lopuszanski:1976vs}
G.~'t~Hooft in {\em Functional and Probabilistic Methods in Quantum Field
  Theory, Vol. 1. Proceedings, XIIth Winter School of Theoretical Physics},
  (Karpacz, Poland), pp.~1318--1322, Feb 17-March 2, 1975.

\bibitem{Boulware:1980av}
D.~G. Boulware, {\it {Gauge Dependence of the Effective Action}},  {\em
  Phys.Rev.} {\bf D23} (1981) 389.

\bibitem{Abbott:1980hw}
L.~Abbott, {\it {The Background Field Method Beyond One Loop}},  {\em
  Nucl.Phys.} {\bf B185} (1981) 189.

\bibitem{Hart:1984jy}
C.~Hart, {\it {Theory and renormalization of the gauge invariant effective
  action}},  {\em Phys.Rev.} {\bf D28} (1983) 1993--2006.

\bibitem{Barvinsky:1985an}
A.~Barvinsky and G.~Vilkovisky, {\it {The Generalized Schwinger-Dewitt
  Technique in Gauge Theories and Quantum Gravity}},  {\em Phys.Rept.} {\bf
  119} (1985) 1--74.

\bibitem{DeWitt:1960fc}
B.~S. DeWitt and R.~W. Brehme, {\it {Radiation damping in a gravitational
  field}},  {\em Annals Phys.} {\bf 9} (1960) 220--259.

\bibitem{Ellicott:1990up}
P.~Ellicott, G.~Kunstatter, and D.~Toms, {\it {Geometrical interpretation of
  the functional measure for supersymmetric gauge theories and of the gauge
  invariant effective action}},  {\em Annals Phys.} {\bf 205} (1991) 70--109.

\bibitem{Rebhan:1987cd}
A.~Rebhan, {\it {Feynman rules and S matrix equivalence of the Vilkovisky-De
  Witt effective action}},  {\em Nucl.Phys.} {\bf B298} (1988) 726.

\bibitem{Rebhan:1986wp}
A.~Rebhan, {\it {The Vilkovisky-De Witt effective action and its application to
  Yang-Mills theories}},  {\em Nucl.Phys.} {\bf B288} (1987) 832.

\bibitem{Fradkin:1983nw}
E.~Fradkin and A.~A. Tseytlin, {\it {On the new definition of off-shell
  effective action}},  {\em Nucl.Phys.} {\bf B234} (1984) 509.

\bibitem{Huggins:1987zw}
S.~Huggins, G.~Kunstatter, H.~Leivo, and D.~Toms, {\it {The Vilkovisky-De Witt
  effective action for quantum gravity}},  {\em Nucl.Phys.} {\bf B301} (1988)
  627.

\bibitem{Pietrykowski}
A.~R. Pietrykowski, ``in preparation.''.

\bibitem{Peeters:2007wn}
K.~Peeters, {\it {Introducing Cadabra: A Symbolic computer algebra system for
  field theory problems}},  \href{http://xxx.lanl.gov/abs/hep-th/0701238}{{\tt
  hep-th/0701238}}.

\bibitem{Peeters:CPC}
K.~Peeters, {\it Cadabra: a field-theory motivated symbolic computer algebra
  system},  {\em Computer Physics Communications} {\bf 176} (2007) 550--558,
  [\href{http://xxx.lanl.gov/abs/cs/0608005v2}{{\tt cs/0608005v2}}].

\bibitem{Gryzov:1992em}
A.~Y. Gryzov, Yu. V.~Kamenshchik and I.~P. Karmazin, {\it {One-loop divergences
  of the einstein theory with a nonminimally interacting scalar field}},  {\em
  Russ.Phys.J} {\bf 35} (1992) 201--205.

\bibitem{Grisaru:1975ei}
M.~T. Grisaru, P.~van Nieuwenhuizen, and C.~Wu, {\it {Background Field Method
  Versus Normal Field Theory in Explicit Examples: One Loop Divergences in S
  Matrix and Green's Functions for Yang-Mills and Gravitational Fields}},  {\em
  Phys.Rev.} {\bf D12} (1975) 3203.

\bibitem{Donkin:2012ud}
I.~Donkin and J.~M. Pawlowski, {\it {The phase diagram of quantum gravity from
  diffeomorphism-invariant RG-flows}},
  \href{http://xxx.lanl.gov/abs/1203.4207}{{\tt arXiv:1203.4207}}.

\bibitem{abramowitz+stegun}
M.~Abramowitz and I.~A. Stegun, {\em Handbook of Mathematical Functions with
  Formulas, Graphs, and Mathematical Tables}.
\newblock Dover, New York, ninth dover printing, tenth gpo printing~ed., 1964.

\bibitem{Narain:2009gb}
G.~Narain and C.~Rahmede, {\it {Renormalization Group Flow in Scalar-Tensor
  Theories. II}},  {\em Class.Quant.Grav.} {\bf 27} (2010) 075002,
  [\href{http://xxx.lanl.gov/abs/0911.0394}{{\tt arXiv:0911.0394}}].

\end{thebibliography}
\end{document}